\documentclass[11pt]{amsart}
\usepackage{amsmath,amssymb}
\usepackage{graphicx}
\usepackage[all]{xy}
\newcommand{\assign}{:=}
\newcommand{\aff}[1]{\underline{#1}}
\newcommand{\tmem}[1]{{\em #1\/}}
\newcommand{\tmmathbf}[1]{\ensuremath{\boldsymbol{#1}}}

\newcommand{\tmop}[1]{\ensuremath{\operatorname{#1}}}

\newenvironment{itemizedot}{\begin{itemize}
    }{\end{itemize}}

\newtheorem{definition}{Definition}[section]
\newtheorem{corollary}[definition]{Corollary}
\newtheorem{theorem}[definition]{Theorem}
\newtheorem{example}[definition]{Example}
\newtheorem{lemma}[definition]{Lemma}
\newtheorem{proposition}[definition]{Proposition}
\newtheorem{remark}[definition]{Remark}

\theoremstyle{plain}
\theoremstyle{definition}

\def\Gr{\mathcal{G}r}
\def\Ic{\mathfrak{I}}
\def\Jc{\mathfrak{J}}

\def\hb{\mathbf{h}}
\def\xb{\mathbf{x}}
\def\ub{\mathbf{u}}

\def\xbb{\underline{\xb}}

\def\zb{\mathbf{z}}
\def\Ac{\mathcal{A}}
\def\NN{\mathbb{N}}
\def\kk{\mathbb{K}}
\def\PP{\mathbb{P}}
\def\Hb{\mathbf{H}}

\def\HilbPPn{\mathrm{Hilb}^{\mu}(\PP^{n})}
\def\Hilb{\mathbf{Hilb}^{\mu}_{\PP^{n}}}
\def\Gr{\mathbf{Gr}}

\def\<{\langle}
\def\>{\rangle}

\title{The Hilbert scheme of points and its link with border basis}

\author{M.E. Alonso$^{(1)}$}
\thanks{$^{(1)}$ Partially supported by, Spanish MEC Sab-PR2007-0133, and
  MTM2008-00272, and by CCG07-UCM-2160} 
\address{M.E. Alonso,  Departmento de Algebra, UCM, 28040 Madrid, Spain}
\email{mariemi@mat.ucm.es}

\author{J. Brachat}
\address{J. Brachat, GALAAD, INRIA, BP 93, 06902 Sophia Antipolis, France}
\email{jbrachat@sophia.inria.fr}

\author{B. Mourrain}
\address{B. Mourrain, GALAAD, INRIA, BP 93, 06902 Sophia Antipolis, France}
\email{mourrain@sophia.inria.fr}

\begin{document}

\maketitle

\begin{abstract}
This paper examines the effective representation of the punctual Hilbert
scheme. We give new equations, which are simpler than Bayer and
Iarrobino-Kanev equations. These new Pl\"ucker-like equations define the
Hilbert scheme as a subscheme of a single Grassmannian and are of degree two
in the Pl\"ucker coordinates.  This explicit complete set of defining
equations for $\HilbPPn$ are deduced from the commutation relations
characterising border bases and from generating equations. We also prove that
the punctual Hilbert functor $\Hilb$ can be represented by the scheme
$\HilbPPn$ defined by these relations and the well-known Pl\"ucker relations
on the Grassmanian.  A new description of the tangent space at a point of the
Hilbert scheme, seen as a subvariety of the Grassmannian, is also given in
terms of projections with respect to the underlying border basis.
\end{abstract}

\section{Introduction}

A natural question when studying systems of polynomial equations is how to
characterize the family of ideals which defines a fixed number $\mu$ of
points counted with multiplicities.  It is motivated by practical issues
related to the solution of polynomial systems, given with approximate
coefficients. Understanding the allowed deformations of a zero-dimensional
algebra, which keep the number of solutions constant, is an actual challenge,
in the quest for efficient and stable numerical polynomial solvers. From a
theoretical point of view, this question is related to the study of the
Hilbert Scheme of $\mu$ points, which is an active area of investigation in
Algebraic Geometry.

The notion of Hilbert Scheme was introduced by \cite{Grothendieck61}: it is
defined as a scheme representing a contravariant functor from the category of
schemes to the one of sets. This functor associates to any scheme $S$ the set
of flat families $\chi \subset \mathbb{P}^{r} \times S$ of closed subschemes
of $\mathbb{P}^{r}$ parametrized by $S$, whose fibers have Hilbert polynomial
$\mu$. 

Many works were developed to analyze its geometric properties (see
eg. \cite{MR927986}), which are still not completely understood. Among them,
it is known to be reducible for $n>2$ \cite{MR0301010}, but the components
are not known for $\mu \geq 8$ \cite{cartwright-2008}. Its connectivity
firstly proved by Hartshorne (1965), is studied in \cite{MR2123227} with a
more constructive approach. 

Though the Hilbert functor $\Hilb$ is known to be representable
\cite{sernesi}, its effective representation is under investigation.  Using
the persistence theorem of \cite{Gotzmann}, a global explicit description of
the Hilbert scheme as a subscheme of a product of two Grassmannians is given
in \cite{MR1735271}, and in \cite{MR2073194} for a multi-graded extension.
Equations defining $\HilbPPn$ in a single Grassmannian are also given in
\cite{MR1735271}. These equations, obtained from rank conditions in the
vector space of polynomials in successive ``degrees'', have a high degree in
the Pl\"ucker coordinnates, namely the number of monomials of degree $\mu$ in
$n+1$ variables minus $\mu$.

In \cite{D.Bayer:82}, a different set of equations of degree $n$ in the
Pl\"ucker coordinates is proposed. It is conjectured that these equations
define the Hilbert scheme, which is proved in \cite{MR2073194}. Nevertheless,
these equations are not optimal, as noticed in an example in dimension
$3$ in \cite{MR2073194}[p. 756]: they are of degree $3$ whereas the
corresponding Hilbert scheme can be defined  in this case by quadratic
equations.

The problem of representation is also studied through subfunctor
constructions and open covering of charts of the Hilbert scheme. Covering
charts corresponding to subsets of ideals with a fixed initial ideal for a
given term ordering are analysed in several works, starting with
\cite{CarraFerro88}, and including more recent one like \cite{lella-2009}. These
open subsets can be embedded into affine open subsets of the Hilbert scheme,
corresponding to ideals associated to quotient algebras with a given monomial
basis. Explicit equations of these affine varieties are developped in
\cite{MR1661369} for the planar case, \cite{MR1905194}, \cite{Huibregtse06},
using syzygies or in \cite{roggero-2009}.

In this paper, we concentrate on the Hilbert scheme of $\mu$ points in the
projective space $\PP_{\kk}^{n}$ and on its effective representation. We give
new equations for the punctual Hilbert scheme, which are simpler than Bayer
and Iarrobino-Kanev equations.  They are quadratic in the Pl\"ucker
coordinnates, and define the Hilbert scheme as a subscheme of a single
Grassmannian. We give a new proof that Hilbert functor can be represented by
this scheme $\HilbPPn$ given by this explicit quadratic
equations. Reformulating a result in \cite{BMnf99}, we recall how the open
chart corresponding to quotient algebras with fixed (monomial) basis
connected to 1 can simply be defined by the commutation relations
characterising border basis (see also \cite{robbiano-2008},
\cite{kr:08:cm}). We show how these commutation relations can be further
exploited to provide an explicit complete set of defining equations for
$\HilbPPn$ as a projective variety.  Following a dual point of view, this
approach yields new Pl\"ucker-like equations of degree two in the coordinates
on the Grassmannian, which are explicit and of smaller degree than those in
\cite{D.Bayer:82}, \cite{MR1735271}. We show moreover that the scheme
$\HilbPPn$ defined by these relations and the Pl\"ucker relations on the
Grassmanian represents the punctual Hilbert functor $\Hilb$. It has a
natural structure of projective variety, as a subvariety of the Grassmannian.
Finally, we give a new description of the tangent space to this variety in
terms of projections with respect to the underlying border basis.

After setting our notations, we analyse in section 2 the Hilbert functor,
starting with a local description based on commutation relations, followed by
and subfunctor constructions. In Section 3, we describe the new
quadratic equations in the Pl\"ucker coordinnates, related them with the
commutation and generating relation for border basis, prove that they
characterise completely elements of the punctual Hilbert scheme and deduce an
explicit representation of the Hilbert functor. Finally in Section 4, we show
how the tangent space to the Hilbert scheme at a given point can be defined
in terms border basis computation. Standard results on functors and on 
genericity are collected in an appendix for the seek of self-contain.

\subsection{Notations}\label{notation}
Let $\kk$ be an algebraically closed field of characteristic 0 and $R = \kk[x_1, \ldots, x_n] = \kk
[\xbb]$ be the set of polynomials in the variables $x_1, \ldots ., x_n$ and
coefficients in $\kk$. We also denote by $S=\kk[x_{0},\ldots,x_{n}]=\kk[\xb]$
the polynomial ring in $x_{0},\ldots,x_{n}$ for a new variable $x_{0}$ "of
homogenization". 
For any $\alpha\in \NN^{n+1}$ (resp. $\NN^{n}$), let $\xb^{\alpha}=
x_{0}^{{\alpha_{0}}}\cdots x_{n}^{\alpha_{n}}$ (resp. $\xbb^{\alpha}=
x_{1}^{\alpha_{1}}\cdots x_{n}^{\alpha_{n}}$). 
The canonical basis of $\NN^{n+1}$ is denoted by $(e_{i})_{i=0, \ldots, n}$, 
so that  $\xb^{\alpha+ e_{i}} = \xb^{\alpha}\, x_{i}$ ($\alpha \in
\NN^{n+1},i=0,\ldots,n$). For $\alpha=(\alpha_{0},\ldots, \alpha_{n})\in \NN^{n+1}$, we denote by 
$\aff{\alpha}= (\alpha_{1},\ldots, \alpha_{n}) \in \NN^{n}$.

For a given set $B=\{ \xb^{\alpha_1}, \ldots, \xb^{\alpha_D}
\}$ of monomials in $x_0, \ldots, x_n$, we  will identify $B$ with
its set of exponents $\{ \alpha_{1},\ldots,\alpha_{D} \} $.
For a set of exponents $E=\{ \alpha_{1},\ldots,\alpha_{D} \} \subset \NN^{n+1}$, 
we denote by $\xb^{E}$ the corresponding set of monomials 
with exponents in $E$: $\{ \xb^{\alpha_1}, \ldots, \xb^{\alpha_D}
\}$.

For $B \subset \NN^{n}$, we say that $B$ is connected to $1$ if $\mathbf{0}
\in B$ (i.e $1 \in \xbb^{B}$)  and
for all $\beta\in B \setminus \{ \mathbf{0} \}$,  there exist $\beta'\in B$ and $i\in
1\ldots n$ such that  $\beta = \beta' + e_{i}$ (i.e $\xbb^{\beta}=
\xbb^{\beta'}\, x_{i}$).

Given an ideal $I$ of $S$, we denote by $I_{d}$ the vector space of homogeneous
polynomials of degree $d$ that belong to $I$. We also denote $s_{d}$ the
dimension of the vector space $S_{d}$ of polynomials $\in S$ of degree $d$. 

For $B \subset \NN^{n}$, we denote by
$B^+ = e_1 + B \cup \cdots\cup e_n  + B \cup B$ and 
$\partial B = B^+ - B$.

A {\em rewriting family} associated to a set $B\subset R$ of monomials is a
set of polynomials of the form $(h_{\alpha})_{\alpha \in \partial B}$ with: 
$$
h_{\alpha}(\xbb) = \xbb^{\alpha} - \sum_{\beta \in B}\, z_{\alpha , \beta }  \, \xbb^{\beta} 
$$
with $z_{\alpha , \beta } \in \kk$ for all $\alpha\in \partial B, \ \beta \in B$.
We call it a {\em border basis} of $B$ if moreover $B$ is a basis of $\Ac=R/(h_{\alpha}(\xbb))$.

If $B =(\beta_{1},\ldots,\beta_{m})$ is a sequence of elements of $\NN^{n}$ and $\beta\in
\NN^{n}$, $B^{\beta_{i}| \beta}$ is the sequence $(\beta_{1},\ldots,\beta_{i-1}$, $\beta,$ $\beta_{i+1},\ldots,\beta_{m})$ obtained
from $B$, by replacing $\beta_{i}$ by $\beta$. Finally we denote by $\langle B \rangle$ the vector space generated by $B$.

We study the set of $\kk$-algebras $\Ac$ generated by $x_{1},\ldots,x_{n}$, that
admit $B$ as a monomial basis. For any $a \in \Ac$, we consider the operator $M_a$ of
multiplication by $a$ in $\Ac$:
\begin{eqnarray*}
  M_a & : & \Ac \rightarrow \Ac\\
  &  & b \mapsto ab
\end{eqnarray*}
As $\mathcal{A}$ is a commutative algebra, the multiplication operators by
the variables $x_i$ commute. Thus for any $p \in R$, we can define the
operator $p (M_{x_1}, \ldots, M_{x_n})$ obtained by substitution of the
variable $x_i$ by $M_{x_i}$ ($i=1,\ldots,n$).

We define $\mathfrak{I} ( \mathcal{A}) \assign
\{p \in R ; p (M_{x_1}, \ldots ., M_{x_n}) = 0\}$ and call it the ideal
associated to $\mathcal{A}$.

The ``dehomogenization by $x_{0}$'' is the application from
$S$ to $R$ that maps a polynomial $p \in S$ to $p(1,x_{1},\ldots,x_{n})
\in R$. For any subset $I \subset S$, we denote $\underline{I}$ its image
by the dehomogenization.\\

Let $A$ be a ring and $I$ an ideal of $A[x_{0},\ldots,x_{n}]$, we will say that a polynomial $P$ is not a zero divisor of $I$ if $I:P=I$.

\section{Hilbert functor representation}
In this section we give a new proof of the existence of the Hilbert scheme $\HilbPPn$ using border basis relations. We will focus on open subfunctors of the Hilbert functor $\mathbf{Hilb}^{\mu}_{\PP^{n}}$ that are represented by affine schemes and consist of a covering of $\HilbPPn$. We will use border bases and commutation relations to define these open affine subschemes of $\HilbPPn$.

\subsection{Border basis representation}\label{sec:bb}
Let $A$ be a local noehterian ring with maximal ideal $m$ and residue field $k:=
A/m$. Suppose that $\mathcal{A}$ is a quotient algebra of $A[x_{1},\ldots,x_{n}]$ that is a
free $A$-module. Assume that $\mathcal{A}$ has a monomial
basis $B$ of size $\mu$, connected to $1$. Then for any $\alpha \in \partial
B$, the monomial $\xbb^{\alpha}$ is a linear 
combination in $\mathcal{A}$ of the monomials of $B$: For any $\alpha \in
\partial B$, there exists $z_{\alpha, \beta} \in \kk$ ($\beta \in B)$
such that \ $h_{\alpha}^{\tmmathbf{z}} ( \xbb) : =
\xbb^{\alpha} - \sum_{\beta \in B} z_{\alpha, \beta}\, 
\xbb^{\beta} \equiv 0$ in $\mathcal{A}$. \ The equations
$h_{\alpha}^{\tmmathbf{z}} ( \xbb)$ will be called, hereafter, the
{\tmem{border relations}} of $\mathcal{A}$ in $B$.

Given these border relations, we define a projection $N^\zb : \langle B^+
\rangle \rightarrow \langle B^+ \rangle$ by:
\begin{itemizedot}
  \item $N^\zb ( \xbb^{\beta}) = \xbb^{\beta}$ if $\beta \in B$,
  
  \item $N^\zb ( \xbb^{\alpha}) = \xbb^{\alpha} - h_{\alpha}^\zb (
  \xbb) = \sum_{\beta \in B} z_{\alpha, \beta} \hspace{0.25em}
  \xbb^{\beta}$ if $\alpha \in \partial B$.
\end{itemizedot}
This construction is extended by linearity to $\langle B^+ \rangle$.

Similarly, the tables of multiplication $M_{x_i}^\zb : \langle B^{} \rangle
\rightarrow \langle B^{} \rangle$ \ are constructed using $M_{x_i}^\zb (
\xbb^{\beta}) = N^\zb (x_i \xbb^{\beta})$ for $\beta \in B$.
Notice that the coefficients of the matrix of $M_{x_i}^{\zb}$ in the
basis $B$ are linear in the coefficients $\zb$.

More generally, a monomial $m$ can be reduced modulo the polynomials
$(h_{\alpha}^{\zb} (\xbb))_{\alpha \in \partial B}$ to a
linear combination of monomials in $B$, as follows: decompose $m = x_{i_1}
\cdots x_{i_l}$ and compute $N^\zb (m) = M_{i_1}^{\tmmathbf{z}} \circ \cdots
\circ M_{i_l}^{\tmmathbf{z}} (1)$. We easily check that $m -
N^\zb (m) \in (h_{\alpha}^{\tmmathbf{z}} ( \xbb))_{\alpha \in \partial
B}$.

Given a free quotient algebra of $A[x_{1},\ldots,x_{n}]$ with basis $B$ connected to 1 of size $\mu$, we have seen that
there exist coefficients $(z_{\alpha, \beta} \in \kk)_{\alpha \in \partial B, \beta \in B}$  which
describe completely an ideal defining $\mu$ points with
multiplicity. Conversely, we are interested in characterizing the coefficients
$\mathbf{z} \assign (z_{\alpha, \beta})_{\alpha \in \partial B, \beta \in B}$ such that the
polynomials $(h_{\alpha}^{\mathbf{z}} ( \xbb))_{\alpha \in B}$ are
the border relations of some quotient algebra $\Ac^{\mathbf{z}}$
in the basis $B$.

The following result \cite{BMnf99}, also
used in {\cite{robbiano-2008, kr:08:cm}} for special cases of base $B$ and adapted to local rings,
answers the question:
\begin{theorem}\label{thm:2.1}
Let $B$ be a set of $\mu$ monomials connected to $1$. The polynomials $h_{\alpha}^{\mathbf{z}} ( \xbb)$ are the border
  relations of some free quotient algebra $\Ac^{\mathbf{z}}$ of $A[x_{1},\ldots,x_{n}]$ of basis $B$ iff
  \begin{equation}
    M_{x_i}^{\mathbf{z}} \circ M_{x_j}^{\mathbf{z}} -
    M_{x_j}^{\mathbf{z}} \circ M_{x_i}^{\mathbf{z}} = 0 \hspace{1em}
    \tmop{for} \hspace{1em} 1 \leqslant i < j \leqslant n. \label{rel:commute}
  \end{equation}
\end{theorem}

\begin{proof}
See {\cite{BMnf99}}.
\end{proof}
In the next propositions, we consider $\tmmathbf{z}=(z_{\alpha, \beta})_{\alpha \in \partial B, \beta \in B}$ as variables.
Then, note that the relations (\ref{rel:commute}) induce polynomial equations of degree $\leqslant 2$ in
$\tmmathbf{z}$ that we will denote:
  \begin{equation}
    M_{x_i}(\tmmathbf{z}) \circ M_{x_j}(\tmmathbf{z}) -
    M_{x_j}(\tmmathbf{z}) \circ M_{x_i}(\tmmathbf{z}) = 0 \hspace{1em}
    \tmop{for} \hspace{1em} 1 \leqslant i < j \leqslant n. \label{eq:commute}
  \end{equation}

%\newpage

\begin{proposition}\label{H}
 Let $B \subset A[x_{1},\ldots,x_{n}]$ be a set of $\mu$ monomials connected to 1. Let $N$ be the size of $\partial B$, then  
$$
\{ I \subset A[x_{1},\ldots,x_{n}]\ |\  \mathcal{A}=R/\underline{I} \ is \ free \ with \ basis\ B \}
$$
is a variety of $\kk^{\mu \times N}$ in the variables $\tmmathbf{z} \in \mathbb{A}_{\kk}^{\mu \times N}$ defined by $\mathfrak{H}_{B} \assign \{ \tmmathbf{z} \in
  \kk^{\mu \times N} ; M_{x_i}(\tmmathbf{z}) \circ M_{x_j}(\tmmathbf{z}) - M_{x_j}(\tmmathbf{z}) \circ M_{x_i}(\tmmathbf{z})
  = 0, 1 \leqslant i < j \leqslant n\}$. We call it the variety of free quotient
  algebras with basis $B$.
\end{proposition}
These varieties depending on monomial sets $B$ are used in \cite{Huibregtse06}
to define the global punctual Hilbert scheme, via a glueing construction.\\
Hereafter, we will give a direct and explicite construction of the Hilbert scheme, based on these relations.\\

\begin{example}\label{ex:mu=2}
To illustrate the construction, we consider the very simple case where
$B=(1,x)$ connected to $1$ in $\kk[x,y]$. Then we have $\partial B =
(y,xy,x^{2})$ and the formal border relations are:
\begin{eqnarray*}
f_{y} &=& y - z_{y,1} + z_{y,x}\,x \\
f_{xy} &=& xy - z_{xy,1} + z_{xy,x}\, x\\
f_{x^{2}} &=& x^{2} - z_{x^{2},1} + z_{x^{2},x}\,x
\end{eqnarray*} 
where $z_{y,1},  z_{y,x},  z_{xy,1}, z_{xy,x}, z_{x^{2},1}, z_{x^{2},x}$ are
the $6$ variables of the border relations. The multiplication matrices are:
$$
M_x = \left( \begin{array}{cc}
     0 & z_{x^{2},1} \\
     1 & z_{x^{2},x} \\
   \end{array} \right),
\ 
M_y = \left( \begin{array}{cc}
     z_{y,1} & z_{xy,1} \\
     z_{y,x} & z_{xy,x} \\
   \end{array} \right).
$$
The equations of $\mathfrak{H}_{B}$ are then given by $M_x\,M_y - M_y\,M_x = 0$.
This yields the following equations of degree 2 in the 6 variables $(z_{\alpha,\beta})$:
 $$ 
\left\{\begin{array}{l}
{z_{xy,1}}-{z_{x^{2},1}}\,{z_{y,x}}=0,\\
{z_{xy,x}}-{z_{y,1}}-{z_{y,x}}\,{z_{x^{2},x}}=0, \\
{z_{x^{2},1}}\,{z_{y,1}}+{z_{x^{2},x}}\,{z_{xy,1}}-{\it z_{xy,x}}\,{z_{x^{2},1}}=0,\\
{z_{x^{2},1}}\,{z_{y,x}}-{z_{xy,1}}=0
\end{array}
\right.
$$
defining the ideal generated by the two polynomials ${z_{xy,1}}-{z_{x^{2},1}}\,{z_{y,x}},
{z_{xy,x}}-{z_{y,1}}-{z_{y,x}}\,{z_{x^{2},x}}$, which define a (parameterized)
variety of dimension $4$.
\end{example}

\subsection{The Hilbert functor }

Let $\mu \in \mathbb{N}$ and $I$ be and ideal of $S$. 
The {\em Hilbert function} of $I$ associates to 
$k\in \NN$ the dimension of $S_{k}/I_{k}$. It coincides with a polynomial called the
{\em Hilbert polynomial} of $I$, for $k$ large enough.\\
We consider the category $\mathcal{C}$ of noetherian schemes over $\mathbb{K}$. Let $\mathbb{P}^{n}$ be the projective scheme $\mathbf{Proj}(S)$. Let $A$ be a commutative ring and $p$ be a prime of $A$. We will denote by $A_{p}$ the localization of $A$ by $p$. Let $m_{p}$ be its maximal ideal. We will denote by $k(p)$ the residue field $A_{p}/A_{p}m_{p}$.

\begin{definition}
 Let $I$ be a graded ideal of S. I is said to be saturated if for all integers k and d such that k $\leq$ d, $I_{d}:S_{k}=I_{d-k}$.
\end{definition}

\begin{definition}
Let $X$ and $Y$ be schemes and $f: X \rightarrow Y$ be a morphism of schemes. $X$ is said to be flat over $Y$ if $\mathcal{O}_{X}$ is $f$-flat over $Y$ i.e for every $x\in X$, $O_{X,x}$ is a $O_{Y,f(x)}$ flat module (see \cite{hartshorne}[Chap.III, p.254]).
\end{definition}

\begin{definition}\label{defi}
 The Hilbert functor of $\mathbb{P}^{n}$ relative to $\mu$ denoted $\mathbf{Hilb}^{\mu}_{\PP^{n}}$ is the contravariant functor from the category $\mathcal{C}$ to the category of Sets which maps an object $X$ of $\mathcal{C}$ to the set of flat families  $Z \subset X \times \mathbb{P}^{n}$  of closed subschemes of  $\mathbb{P}^{n}$ parametrized by  $X$ with fibers having Hilbert polynomial  $\mu$ (flat families $Z \subset X \times \mathbb{P}^{n}$ means that $Z$ is flat over $X$).
\end{definition}

\begin{example}\label{ex} 
\rm
If $X = \mathbf{Spec}(A)$, where $A$ is a noetherian $\mathbb{K}$-algebra,
$\mathbf{Hilb}^{\mu}_{\mathbb{P}^{n}}(X)  $ is given by the set of saturated
homogeneous ideals $I$ of $A[x_{0},\ldots,x_{n}]$ such that
$\mathbf{Proj}(A[x_{0},\ldots,x_{n}] / I)$ is flat over $\mathbf{Spec}(A)$
and for every prime ideal $p \subset A$, the Hilbert polynomial of the
$k(p)$-graded algebra $(A[x_{0},\ldots,x_{n}]/I) \otimes_{A} k(p)$ is equal
to $\mu$ where $k(p)$ is the residue field $A_{p}/pA_{p}$.  
\end{example}

\begin{definition}\label{I(p)}
 Let $A$ be a noetherian $\mathbb{K}$-algebra. Let $p \in \mathbf{Spec}(A)$
 be a prime of $A$ with residue field $k(p) := A_{p}/pA_{p}$. Let $I$ be a
 homogeneous ideal of $A[x_{0},\ldots,x_{n}]$. Consider the following exact
 sequence: 
$$
\xymatrix{
0 \ar[r] & I \ar[r] & A[x_{0},\ldots,x_{n}] \ar[r] & A[x_{0},\ldots,x_{n}]/I \ar[r] & 0
}
$$
Tensoring by $k(p)$ we get the exact sequence
$$
\xymatrix{
 I \otimes k(p) \ar[r] & k(p)[x_{0},\ldots,x_{n}] \ar[r] & A[x_{0},\ldots,x_{n}]/I \otimes k(p) \ar[r] & 0
}
$$
Then, we will denote by $I(p)$ the homogeneous ideal of $k(p)[x_{0},\ldots,x_{n}]$ which consists of the image of $I \otimes k(p)$ in $k(p)[x_{0},\ldots,x_{n}]$. Thus we have
$$
A[x_{0},\ldots,x_{n}]/I \otimes k(p) \sim k(p)[x_{0},\ldots,x_{n}]/I(p).
$$
\end{definition}
\begin{remark}
Note that in general $I(p)$ is not isomorphic to $I \otimes k(p)$ because tensoring by $k(p)$ is not a left exact functor i.e the morphism:
$$
\xymatrix{
 I \otimes k(p) \ar[r] & k(p)[x_{0},\ldots,x_{n}]
}
$$
is surjective but not injective.
\end{remark}

In our analysis, we will use the affine setting described in Section \ref{sec:bb}.
In order to identify the good affinizations which lead to this
setting, we introduce the following definition and characterization:
\begin{definition}\label{defpoint}
Given an homogeneous ideal $J$ in
$\mathbf{Hilb}^{\mu}_{\mathbb{P}^{n}}(\overline{k})$, one has from the
Nullstellensatz theorem that $J$ has the following primary decomposition: 
$$
J = \bigcap_{i} q_{i}
$$
with $q_{i}$ homogeneous $m_{\overline{k},P_{i}}$-primary ideal, for some
points $P_{i}$ in the projective space $\mathbb{P}^{n}_{\overline{k}}$. The
set $\{P_{i}\}$ will be called the set of points defined by $J$ in
$\mathbb{P}^{n}_{\overline{k}}$.

More generaly, let $J$ be an homogeneous ideal (not necessarily saturated) of
$S$ with Hilbert polynomial equal to the constant $\mu$. The set of points
defined by $J$  in $\mathbb{P}^{n}_{\overline{k}}$ is the set of points
defined below by its saturation (denoted $Sat(J)$): 
$$
Sat(J):=\bigcup_{j \in \mathbb{N}} J:(m_{\overline{k}})^{j}
$$
in $\mathbb{P}^{n}_{\overline{k}}$.
\end{definition}

\begin{proposition}\label{plat}
 Let $X = \mathbf{spec}(A)$ be a scheme in $\mathcal{C}$ and $Z =
 \mathbf{Proj}(A[x_{0},$ $\ldots, x_{n}]/I)$ be an element of
 $\mathbf{Hilb}^{\mu}_{\mathbb{P}^{n}}(X)$. Let $u$ be a linear form in
 $\mathbb{K}[x_{0},\ldots, x_{n}]$ and $Z_{u}$ be the open set associated to
 $u$ considered as an element of $H^{0}(Z,O_{Z}(1))$ (see
 \cite{EGA1}[(0.5.5.2), p.53]). Let $\pi$ be the natural morphism from $Z$ to
 $X$. Let $p \in \mathbf{Spec}(A)$ be a prime of $A$ and $k(p):=
 A_{p}/pA_{p}$ its residue field. Then, $\pi_{*}(O_{Z_{u}})_{p}$ is a free
 $\mathcal{O}_{X , p}$-module of rank  $\mu$ if and only if $u$ does not
 vanish at any of the points defined by $\overline{I(p)}= I(p)\otimes_{k(p)}
 \overline{k(p)}$ in $\mathbb{P}^{n}_{\overline{k(p)}}$ (see definition
 \ref{defpoint} and \ref{I(p)}), with $\overline{k(p)}$ the
 algebraic closure of $k(p)$.
\end{proposition}
\begin{proof}

By a change of variables in $\mathbb{K}[x_{0},\ldots,x_{n}]$ we can assume
that $\ub = x_{0}$. Moreover, without loss of generality, we can assume that
$A$ is a local ring with maximal ideal $p$.\\ 
Let $\underline{I} \subset A[x_{1},\ldots,x_{n}]$ be the affinization of $I$ by $x_{0}$ (set $x_{0} = 1$).  One has that  $\underline{I}(p) =  \underline{I(p)}$ and that
$$
(A[x_{0},\ldots,x_{n}]/I) \otimes_{A} k(p) = k(p)[x_{0},\ldots,x_{n}]/I(p)
$$
and 
$$
(A[x_{1},\ldots,x_{n}]/\underline{I}) \otimes_{A} k(p) = k(p)[x_{1},\ldots,x_{n}]/\underline{I}(p).
$$
We also know that $Z_{x_{0}} = D_{+}(x_{0})$ (see \cite{EGA2}[Prop (2.6.3), p.37]) and that $Z_{|D_{+}(x_{0})} = \mathbf{Spec}(A[x_{1},\ldots,x_{n}]/\underline{I})$. Thus $\pi_{*}(\mathcal{O}_{Z \ |D_{+}(x_{0})})$ is the sheaf of $\mathcal{O}_{X}$-module associated to the $A$-module $A[x_{1},\ldots,x_{n}]/\underline{I}$ on $\mathbf{Spec}(A)$. Finally, $\pi_{*}(O_{Z_{x_{0}}})_{p}$ is a free $\mathcal{O}_{X , p}$-module of rank  $\mu$ if and only if $A[x_{1},\ldots,x_{n}]/\underline{I}$ is free of rank $\mu$.\\

First let us prove that there exists an integer $N>0$ such that for all $d \geq N$, the multiplication by $x_{0}$:
\begin{equation}
\xymatrix{
k(p)[x_{0},\ldots,x_{n}]_{d}/I(p)_{d} \ar[r]^{*x_{0}} & k(p)[x_{0},\ldots,x_{n}]_{d+1}/I(p)_{d+1}
}
\end{equation}
is injective if and only if $x_{0}$ does not vanish at any of the points defined by $\overline{I(p)}$ in $\mathbb{P}^{n}_{\overline{k(p)}}$.\\
As a matter of fact, $I(p)$ defines $\mu$ points in $k(p)[x_{0},\ldots,x_{n}]$, there exists and integer $N \geq \mu$ such that the dimension of $k(p)[x_{0},\ldots,x_{n}]_{d}/I(p)_{d}$ is equal to $\mu$ for all degree $d \geq N$. One has from \cite{MR1735271}[C.28] that $I(p)_{d+1}:S_{1} = I(p)_{d}$ for all $d \geq N$ i.e:
$$
Sat(I(p)) = \sum_{1 \leq i \leq d}I(p)_{d}:S_{i} + (I(p)_{d}) \ \forall d \geq N.
$$
with $Sat(I(p)):=\bigcup_{d\in\mathbb{N}}I(p):(S_{d})$ (i.e $I(p)$ is saturated in degree greater than $N$).\\
Thus the multiplication by $x_{0}$ is injective (and by dimension bijective) for all $d \geq N$ if and only if $x_{0}$ is not a zero divisor of the saturation $Sat(I(p))$ of $I(p)$. By proposition \ref{vanish} and definition \ref{defpoint}, this is equivalent to $x_{0}$ does not vanish at any of the points defined by $\overline{I(p)}=\overline{k(p)} \otimes I(p)$ in $\mathbb{P}^{n}_{\overline{k(p)}}$.\\

Consider now the following commutative diagram for all $d \geq N$:

\begin{equation}\label{diagram}
\xymatrix{ 
   (A[x_{0},\ldots,x_{n}]_{d}/I_{d})\otimes k(p) \ar[r]^{\sim} \ar@{->>}[d]^{\phi} & k(p)[x_{0},\ldots,x_{n}]_{d}/I(p)_{d} \ar@{->>}[d]^{\delta} \\
   (A[x_{1},\ldots,x_{n}]_{\leq d}/\underline{I}_{\leq d})\otimes k(p) \ar@{->>}[r]^{\psi} \ar[d]^{j} &  k(p)[x_{1},\ldots,x_{n}]_{\leq d}/\underline{I}(p)_{\leq d} \ar[d]^{i} \\
   (A[x_{1},\ldots,x_{n}]/\underline{I})\otimes k(p) \ar[r]^{=} & k(p)[x_{1},\ldots,x_{n}]/\underline{I}(p)
}
\end{equation}

First, as $k(p)[x_{1},\ldots,x_{n}]/\underline{I}(p)$ is a $k(p)$-algebra of dimension less or equal to $\mu$, one has that
\begin{equation}
 \xymatrix{
k(p)[x_{1},\ldots,x_{n}]_{\leq d}/\underline{I}(p)_{\leq d} \ar[r]^{i} & k(p)[x_{1},\ldots,x_{n}]/\underline{I}(p)
}
\end{equation}
is an isomorphism for all $d \geq N \geq \mu$ (this comes from the fact that the Hilbert function of $k(p)[x_{1},\ldots,x_{n}]/\underline{I}(p)$
is strictly increasing until it is the constant function equal to the dimension of $k(p)[x_{1},\ldots,x_{n}]/\underline{I}(p) \leq \mu$  ). Thus, for all $d \geq N$, $j$ is surjective.\\

Let us prove the equivalence between $A[x_{1},\ldots,x_{n}]/\underline{I}$ is a free  $A$-module of rank $\mu$ and $x_{0}$ does not vanish at any point defined by $\overline{I(p)}$ in $\mathbb{P}^{n}_{\overline{k(p)}}$.\\

First, assume $x_{0}$ does not vanish at any point defined by $\overline{I(p)}$ in $\mathbb{P}^{n}_{\overline{k(p)}}$. Thus, the multiplication by $x_{0}$
\begin{equation}\label{multiplix}
\xymatrix{
k(p)[x_{0},\ldots,x_{n}]_{d}/I(p)_{d} \ar[r]^{ x_{0}} & k(p)[x_{0},\ldots,x_{n}]_{d+1}/I(p)_{d+1}
}
\end{equation}
is a bijection for all $d \geq N$ i.e the morphism $\delta$ in diagram (\ref{diagram}) is an isomorphism. Thus all the morphisms in diagram  (\ref{diagram}) are isomorphisms for all $d \geq N \geq \mu$. Consequently, using the multiplication \eqref{multiplix} from degree $d$ to $d+1$, one has that the natural morphism
\begin{equation}
\xymatrix{
(A[x_{1},\ldots,x_{n}]_{\leq d}/\underline{I}_{\leq d})\otimes k(p) \ar[r] & (A[x_{1},\ldots,x_{n}]_{\leq d+1}/\underline{I}_{\leq d+1})\otimes k(p)
}
\end{equation}
is an isomorphism for all $d \geq N$. By Nakayama Lemma, the natural inclusion:
\begin{equation}
\xymatrix{
A[x_{1},\ldots,x_{n}]_{\leq d}/\underline{I}_{\leq d} \ar[r] & A[x_{1},\ldots,x_{n}]_{\leq d+1}/\underline{I}_{\leq d+1}
}
\end{equation}
is an isomorphism and
\begin{equation}\label{neuf}
A[x_{1},\ldots,x_{n}]_{\leq d}/\underline{I}_{\leq d}  =  A[x_{1},\ldots,x_{n}]_{\leq d+1}/\underline{I}_{\leq d+1} = \ldots = A[x_{1},\ldots,x_{n}]/\underline{I}.
\end{equation}
Finally,  $A[x_{1},\ldots,x_{n}]/\underline{I}$ is a flat $A$-module of finite type such that $A[x_{1},\ldots,x_{n}]/\underline{I} \otimes k(p)$ is of dimension $\mu$. Using \cite{ueno}[lem.7.51, p.55], we deduce that $A[x_{1},\ldots,x_{n}]/\underline{I}$ is a free $A$-module of rank $\mu$.\\

Reciprocally, assume that $A[x_{1},\ldots,x_{n}]/\underline{I}$ is a free
$A$-module of rank $\mu$. Then the dimension of
$A[x_{1},\ldots,x_{n}]/\underline{I} \otimes k(p)$ is equal to $\mu$ . Thus
all the morphisms in diagram (\ref{diagram}) (in particular $\delta$) are
isomorphisms for all $d \geq N$. Consequently $x_{0}$ is not a zero divisor
of $Sat(I(p))$ and thus does not vanish at any point defined by
$\overline{I(p)}$ in $\mathbb{P}^{n}_{\overline{k(p)}}$. 
\end{proof}

\begin{example}
Let $A=\kk[s,t]/(s\,t)$ and $I=(s\, x_{0}^{2}+ x_{0}x_{1} + t\,
x_{1}^{2})\subset A[x_{0},x_{1}]$. For each prime ideal $p$ of $A$, the
quotient $k(p)[x_{0},x_{1}]/I$ is of dimension $2$. Thus, $I\in
\mathbf{Hilb}^{2}_{\PP^{1}}(A)$. If we take the prime ideal $p=(s)$ of $A$,
then $A_{p}=k(p)=\kk(t)$ and the roots of $I(p)$ in $\overline{\kk(t)}$ are
$(1:0), (-t:1) \in \PP^{1}(\overline{\kk(t)})$. If we take $u=x_{0}+x_{1}$,
it does not vanish at the roots of $I(p)$. By a change of variables,
$X_{0}=x_{0}+x_{1}, X_{1}=x_{1}$ and taking $X_{0}=1$, we obtain
$\underline{I}=(s + (1+2\,s) \, X_{1} + (1+s+t) X_{1}^{2})$. Thus 
$$
\pi_{*}(O_{Z_{X_{0}}})_{p}=A_{p}[X_{1}]/ \underline{I}_{p} = A_{p}[X_{1}]/(X_{1} + (1+t)
X_{1}^{2})
$$ 
is a free $A_{p}$ module of rank $2$ generated by  $\{1,X_{1}\}$, since $(1+t)$ is
invertible in $A_{p}$. 

If we take $p=(s,t)$, then $k(p)=\kk$ and the roots of $I(p)$ are 
$(1:0), (0:1) \in \PP^{1}(\overline{\kk})$. By the same change of variables, 
we obtain 
$$
\pi_{*}(O_{Z_{X_{0}}})_{p}=A_{p}[X_{1}]/ \underline{I}_{p} = A_{p}[X_{1}]/(s + (1+2\,s) \, X_{1} + (1+s+t) X_{1}^{2})
$$ 
which is also a free $A_{p}$ module of rank $2$ generated by  $\{1,X_{1}\}$, since $(1+s+t)$ is
invertible in $A_{p}$.

Notice that the localisation is needed: $A[X_{1}]/(X_{1} + (1+t)
X_{1}^{2})$ or $A[X_{1}]/(s + (1+2\,s) \, X_{1} + (1+s+t) X_{1}^{2})$
are not free $A$-modules nor of finite type.
\end{example}

\begin{corollary}\label{laremark}
 Let $A$ be a local ring with maximal ideal $m$ and a noetherian $\mathbb{K}$-algebra. Then
$$
\mathbf{Hilb}^{\mu}_{\PP^{n}}(A) = 
$$
$$
\{ \text{Saturated ideal } I \subset A[x_{0},\ldots,x_{n}] | \  A[x_{0},\ldots,x_{n}]_{d}/I_{d} \ \text{is free of rank } \mu \text{ for } d\geq \mu \}.
$$
\end{corollary}
\begin{proof}
First, let $I \subset A[x_{0},\ldots,x_{n}]$ be a saturated homogeneous ideal such that
$A[x_{0},\ldots,x_{n}]_{d}/I_{d}$ is free of rank $\mu$ for all degree $d\geq \mu$, then $I$ belongs to $\mathbf{Hilb}^{\mu}_{\PP^{n}}(A)$.

We deduce $\underline{I}(p)$ defines an affine zero dimensional algebra of
multiplicity $\mu$ in $k(p)[x_{1},\ldots,x_{n}]$. Thus, using homogenization
by $x_{0}$ on $\underline{I}(p)$ we obtain an homogeneous ideal in
$\mathbf{Hilb}^{\mu}_{\mathbb{P}^{n}}(k(p))$. Then, by the Gotzmann's
persistence theorem  \cite{MR1735271}[C.17, p.297] we deduce that for all
degree $d \geq \mu$ the natural inclusion 
$$
i: k(p)[x_{1},\ldots,x_{n}]_{\leq d}/\underline{I}(p)_{\leq d} \longrightarrow k(p)[x_{1},\ldots,x_{n}]/\underline{I}(p)
$$
is an isomorphism (by dimension). Thus, the morphism 
$$
j : (A[x_{1},\ldots,x_{n}]_{\leq d}/\underline{I}_{\leq d})\otimes k(p) \longrightarrow (A[x_{1},\ldots,x_{n}]/\underline{I})\otimes k(p)
$$
is surjective for all $d \geq \mu$. Then, by Nakayama's Lemma, we get that
$$
A[x_{1},\ldots,x_{n}]_{\leq \mu}/\underline{I}_{\leq \mu} = A[x_{1},\ldots,x_{n}]_{\leq \mu+1}/\underline{I}_{\leq \mu+1} = \cdots = A[x_{1},\ldots,x_{n}]/\underline{I}.
$$
Finally, as $x_{0}$ is not a zero divisor of $I$, we get that
$A[x_{1},\ldots,x_{n}]_{\leq d}/\underline{I}_{\leq d} \simeq A[x_{0},\ldots, x_{n}]_{d}/I_{d}$ is a free $A$-module of rank $\mu$ for all degree $d \geq \mu$. 
\end{proof}

\begin{corollary}\label{regularite}
Let $A$ be a local ring with maximal ideal $m$ and a noethrian $\mathbb{K}$-algebra. Let $I \subset A[x_{0},\ldots,x_{n}]$ be a homogeneous ideal in $\mathbf{Hilb}^{\mu}_{\PP^{n}}(A)$. Then $I$ is generated in degree $\mu$:
$$
I_{\mu+k} = A[x_{0},\ldots,x_{n}]_{k}\,I_{\mu}
$$
for all $k \geq 0$.
\end{corollary}
\begin{proof}
As in the proof of corollary \ref{laremark}, we can assume $x_{0}$ is not a zero divisor of $I$ and does not vanish at any point defined by $\overline{I(m)}$ (see definitions \ref{I(p)} and \ref{defpoint}). Then, from proposition \ref{plat}, $A[x_{1},\ldots x_{n}]/\underline{I}$ is a free $A$-module of rank $\mu$ and we get the following diargam:
$$
\xymatrix{ 
   (A[x_{1},\ldots,x_{n}]_{\leq d}/\underline{I}_{\leq d})\otimes k(p) \ar@{->>}[r]^{\psi} \ar[d]^{j} &  k(p)[x_{1},\ldots,x_{n}]_{\leq d}/\underline{I}(p)_{\leq d} \ar[d]^{i} \\
   (A[x_{1},\ldots,x_{n}]/\underline{I})\otimes k(p) \ar[r]^{\sim} & k(p)[x_{1},\ldots,x_{n}]/\underline{I}(p)
}
$$
for which we proved in corollary \ref{laremark} that $i$, $j$ (and thus
$\psi$) are isomorphism for all $d \geq \mu$. As
$k(p)[x_{1},\ldots,x_{n}]/\underline{I}(p)$ is of dimension $\mu$, we can
find a basis $B$ with polynomials of degree less or equal to $\mu -1$ (take
for instance $B$ connected to 1). Thus, $B$ is a basis of
$k(p)[x_{1},\ldots,x_{n}]_{\leq d}/\underline{I}(p)_{\leq d} \sim
(A[x_{1},\ldots,x_{n}]_{\leq d}/\underline{I}_{\leq d})\otimes k(p)$ for all
$d \geq \mu$. By Nakayama's lemma, we deduce that $B$ is a basis of the free
$A$-module $A[x_{1},\ldots,x_{n}]_{\leq d}/\underline{I}_{\leq d}$ for all $d
\geq \mu$. As the degree of all the polynomials in $B$ is less or equal to
$\mu-1 < \mu \leq d$, we can define operators of multiplication by the
variables $(x_{i})_{1 \leq i \leq n}$ in $A[x_{1},\ldots,x_{n}]_{\leq
  d}/\underline{I}_{\leq d}$ for all $d \geq \mu$. Then, using these
operators of multiplication, we can easily prove that $\underline{I}_{\leq
  d+1}  
 = A[x_{1},\ldots,x_{n}]_{\leq 1}.\underline{I}_{\leq d}$ for all $d \geq \mu$. As $x_{0}$ is not a zero divisor of $I$, we deduce
$$
I_{d+1} = S_{1}\, I_{d}
$$
for all $d \geq \mu$.
\end{proof}

\begin{remark}
Let $A$ be a noetherian $\mathbb{K}$-algebra. Let $X = \mathbf{Spec}(A)$ and $Z$ be a closed subscheme of $X \times \mathbb{P}^{n}$ and let $\mathcal{I} \subset \mathcal{O}_{X \times \mathbb{P}^{n}}$ be the sheaf of ideals that defines $Z$.   Corollary \ref{regularite} means that if $Z$ belongs to $\mathbf{Hilb}^{\mu}_{\mathbb{P}^{n}}(X)$, then the natural map
$$
H^{0}(X \times \mathbb{P}^{n},\mathcal{I}(d)) \otimes_{\mathbb{K}} \mathcal{O}_{X \times \mathbb{P}^{n}}(1) \longrightarrow H^{0}(X \times \mathbb{P}^{n},\mathcal{I}(d+1))
$$
is surjective.
\end{remark}

\subsection{Open covering of the Hilbert functor}

Let $A$ be a ring and $M$ is an $A$-module. We denote by $\widetilde{M}$ the
quasi-coherent sheaf of modules associated to $M$ in $\mathbf{Spec}(A)$. We
will say that $M$ is locally free on $\Omega \subset \mathbf{Spec}(A)$ if for
all $p \in \Omega$, $M_{p}$ is a free $A_{p}$-module. We will say that $M$ is
locally free if it is locally free on $\mathbf{Spec}(A)$. Thus, $M$ is
locally free if and only if $\widetilde{M}$ is locally free on
$\mathbf{Spec}(A)$ as a sheaf of modules.\\
We recall some definitions about functors (see eg. \cite{sernesi}[Appendix E]):
\begin{definition}\label{representabl}
A contravariant functor $F$ from the category $\mathcal{C}$ to the category
of Sets is representable if there exists an object $Y$ in $\mathcal{C}$ such
that the functor $\mathbf{Hom}(-,Y)$ is isomorphic to the functor $F$. In
particular for every $X$ in $\mathcal{C}$, 
$$
F(X) \simeq Hom(X,Y).
$$
\end{definition}

\begin{definition}
 A contravariant functor $F$ from the category $\mathcal{C}$ to the category
 of Sets is called a sheaf if for every scheme $X$ in $\mathcal{C}$, the
 presheaf of sets on the topological space associated to $X$ given by: 
$$
U \rightarrow F(U)
$$
is a sheaf. Namely, if for all schemes $X$ in $\mathcal{C}$ and for every open covering $\{U_{i}\}$ of $X$, the following is an exact sequences of sets:
$$
0 \rightarrow F(X) \rightarrow \prod_{i} F(U_{i}) \rightarrow  \prod_{i,j} F(U_{i} \cap U_{j})
$$
\end{definition}
Notice that by construction, representable functors are sheaves.

\begin{definition}\label{subfunctor}
Let $F$ be a contravariant functor from $\mathcal{C}$ to the category of
Sets. A subfunctor $G$ of $F$ is said to be an open subfunctor if for every
scheme $X$ in $\mathcal{C}$ and for every morphism of functors 
$$
\mathbf{Hom}(-,X) \rightarrow F
$$ 
the fiber product $\mathbf{Hom}(-,X) \times_{F} G$ (which is a subfunctor of $\mathbf{Hom}(-,X)$) is represented by an open subscheme of $X$.\\
A family of open subfunctors $\{ G_{i} \}$ of $F$ is a covering if for every scheme $X$ in $\mathcal{C}$, the family of subschemes that represent the subfunctors  $\{ \mathbf{Hom}(-,X) \times_{F} G_{i} \}$ is an open covering of $X$.
\end{definition}

\begin{definition}\label{hudef}
 Let $u$ be a linear form in $\mathbb{K}[x_{0},\ldots,x_{n}]$. Let $\Hb_{u}$
 be the subfunctor of $\mathbf{Hilb}^{\mu}_{\PP^{n}}$ which associates to $X$
 in $\mathcal{C}$ the set $\Hb_{u}(X)$ of flat families  $Z \subset X \times
 \mathbb{P}^{n}$  of closed subschemes of  $Y$ parametrized by  $X$ with
 fibers having Hilbert polynomial  $\mu$ and such that $\pi_{*}(O_{Z_{u}})$
 is locally free sheaf of rank $\mu$ of X, where $\pi$ is the natural
 morphism from $Z$ to $X$ and $Z_{u}$ is the open set associated to $u$
 considered as an element of $H^{0}(Z,O_{Z}(1))$ (see \cite{EGA1}[(0.5.5.2) p.53]). 
\end{definition}

\begin{proposition}\label{hu}
 The family of subfunctors $(\Hb_{u})_{u \in \mathbb{K}[x_{0},\ldots,x_{n}]_{1}}$ consists of an open covering of subfunctors of $\mathbf{Hilb}^{\mu}_{\PP^{n}}$. 
\end{proposition}
\begin{proof}
From proposition \ref{functor}, it is enough to consider affine schemes $X=\mathbf{Spec}(A)$ (with $A$ a noetherian $\mathbb{K}$-algebra) and to prove that, given a morphism of functors from $\mathbf{Hom}(-,X)$ to $\mathbf{Hilb}^{\mu}_{\PP^{n}}$  (i.e given an element $Z \in \mathbf{Hilb}^{\mu}_{\mathbb{P}^{n}}(X) $) the functor
$$
G:=\mathbf{Hom}(-,X) \times_{\mathbf{Hilb}^{\mu}_{\PP^{n}}} \Hb_{u} 
$$
restricted to the category of affine noetherian schemes over $\mathbb{K}$ is represented by an open subscheme of $X$.\\
Let $X' = \mathbf{Spec}(A')$ be an affine noetherian scheme over
$\mathbb{K}$. Let $f$ be a morphism of $\mathbb{K}$-algebras from $A$ to
$A'$. Let $\phi$ be the morphism of schemes from $X' = \mathbf{Spec}(A')$ to
$\mathbf{Spec}(A)$ associated to $f$. Let $Z$ be an element of
$\mathbf{Hilb}^{\mu}_{\mathbb{P}^{n}}(X) $ and $I$ be its associated
saturated homogeneous ideal of $A[x_{0},\ldots,x_{n}]$. Let $Z'$ be the
element of $\mathbf{Hilb}^{\mu}_{\PP^{n}}(X')$ given by $(\phi \times
Id_{\mathbb{P}^{n}})^{*}(Z)$. Let $I'$ be the homogeneous ideal of
$A'[x_{0},\ldots,x_{n}]$ associated to the quotient algebra
$(A[x_{0},\ldots,x_{n}]/I) \otimes_{A} A'$.\\ 
By a change of variables in $\mathbb{K}[x_{0},\ldots,x_{n}]$, we can assume
$u=x_{0}$. Then, $Z'_{x_{0}} = (\phi \times
Id_{\mathbb{P}^{n}})^{*}(Z_{x_{0}})$ is equal to
$\mathbf{Spec}(A'[x_{1},\ldots,x_{n}]/\underline{I}')$ (where $\underline{ I
}'$ denote the affinization of $I'$).\\ 
Then, we need to prove that for all $A'$ and $f$,  the sheaf of module of
$\mathbf{Spec}(A')$ associated to the $A'$-module
$A'[x_{1},\ldots,x_{n}]/\underline{I}'$ is locally free of rank $\mu$ if and
only if the morphism  
$$
 \phi: \mathbf{Spec}(A')  \rightarrow \mathbf{Spec}(A) 
$$
factors through an open subscheme $\Omega_{x_{0}}$ of $\mathbf{Spec}(A)$.\\

Let $q$ be a prime of $A'$ and $p$ be its image in $\mathbf{Spec}(A)$. From proposition \ref{plat}, $A'_{q}[x_{1},\ldots,x_{n}]/\underline{I}'_{q}$ is free of rank $\mu$ if and only if  $x_{0}$ does not vanish at any point defined by $\overline{I'(q)} = I'(q) \otimes_{k'(q)} \overline{k'(q)}$ in $\mathbb{P}^{n}_{\overline{k'(q)}}$ (see Definition \ref{I(p)}). Note that we have:
$$
I'(q) = I(p)\otimes_{k(p)} k'(q).
$$
Thus, by Proposition \ref{point}, the points defined by $\overline{I'(q)}$ are the same as those defined by $\overline{I(p)} = I(p) \otimes_{k(p)} \overline{k(p)}$ in $\mathbb{P}^{n}_{\overline{k(p)}}$ using the natural field inclusions
$$
 \xymatrix{ 
    k(p) \ar[r]^{i} \ar[d]  & k'(q) \ar[d] \\
    \overline{k(p)} \ar[r]^{\overline{i}}  &  \overline{k'(q)}
}
$$

Thus, $x_{0}$ does not vanish at the points defined by $\overline{I'(q)}$ if and only if $x_{0}$ does not vanish at the points defined by $\overline{I(p)}$. Equivalently: $A'_{q}[x_{1},\ldots,x_{n}]/\underline{I}_{q}'$ is  free of rank $\mu$ if and only if $A_{p}[x_{1},\ldots,x_{n}]/\underline{I}_{p}$ is  free of rank $\mu$ (with $p = \phi(q)$). Thus, $A'[x_{1},\ldots,x_{n}]/\underline{I}'$ is locally free of rank $\mu$ if and only if $\phi$ factors through the subset $\Omega_{x_{0}} \subset \mathbf{Spec}(A)$ on which $A[x_{1},\ldots,x_{n}]/\underline{I}$ is locally free of rank $\mu$.\\
Let $p \in \mathbf{Spec}(A)$ be a prime of $A$. Then, using diagram \ref{diagram}, one has the following equivalence: \\ 
\ (i) $A[x_{1},\ldots,x_{n}]/\underline{I} \otimes_{A} A_{p} = A_{p}[x_{1},\ldots,x_{n}]/\underline{I_{p}}$ is an $A_{p}$ free module,\\
\ (ii) $A[x_{1},\ldots,x_{n}]/\underline{I} \otimes_{A} A_{p} = A_{p}[x_{1},\ldots,x_{n}]/\underline{I_{p}}$ is an $A_{p}$-module of finite type,\\
\ (iii) $A_{p}[x_{1},\ldots,x_{n}]_{\leq d}/\underline{I_{p}}_{\leq d} = A_{p}[x_{1},\ldots,x_{n}]_{\leq d+1}/\underline{I_{p}}_{\leq d+1}$ for all $d \geq \mu$,\\
\ (iv) there exists an integer $d \geq \mu$ such that 
$$A_{p}[x_{1},\ldots,x_{n}]_{\leq d}/\underline{I_{p}}_{\leq d} = A_{p}[x_{1},\ldots,x_{n}]_{\leq d+1}/\underline{I_{p}}_{\leq d+1}.$$\
\\
Thus, the set $\Omega \subset \mathbf{Spec}(A)$ on which $A[x_{1},\ldots,x_{n}]/\underline{I}$ is locally free is equal to the subset of $\mathbf{Spec}(A)$ on which the morphism of inclusion
$$
i: A[x_{1},\ldots,x_{n}]_{\leq \mu}/\underline{I}_{\leq \mu} \longrightarrow A[x_{1},\ldots,x_{n}]_{\leq \mu+1}/\underline{I}_{\leq \mu+1}
$$
is surjective. Let $M$ be the cokernel of $i$, then $\Omega$ is equal to the subset of $\mathbf{Spec}(A)$ on which $\widetilde{M}$ is equal to zero. As $M$ is of finite type (i.e $\widetilde{M}$ is a coherent scheaf of module), $\Omega$ is an open subset of $\mathbf{Spec}(A)$.\\
Finally, $\Omega_{x_{0}}$ is an open subset of $\Omega$ and $A'[x_{1},\ldots,x_{n}]/\underline{I}'$ is locally free of rank $\mu$ if and only if $\phi$ factors through the open subscheme associated to $\Omega_{x_{0}}$. Consequently, $\Hb_{x_{0}}$ is an open subfunctor of $\mathbf{Hilb}^{\mu}_{\PP^{n}}$.\\

Let us prove that $(\Hb_{u})_{u \in \mathbb{K}[x_{0},\ldots,x_{n}]_{1}}$  consists of a covering of $\mathbf{Hilb}^{\mu}_{\PP^{n}}$. By definition \ref{subfunctor}, we need to prove that $(\Omega_{u})_{u \in S_{1}}$ consists of a covering of $\mathbf{Spec}(A)$. Let $p$ be a point of $\mathbf{Spec}(A)$. Consider the points of $\mathbb{P}^{n}_{k(p)}$ defined by $I(p)$. One can find a linear form $u \in S_{1} = \mathbb{K}[x_{0},\ldots,x_{n}]_{1}$ that does not vanish at any of these points. By proposition \ref{plat}, $p$ belongs to $\Omega_{u}$.\\ 
Thus the family $(\Omega_{u})_{u \in \mathbb{K}[x_{0},\ldots,x_{n}]_{1}}$ consists of a covering of $\mathbf{Spec}(A)$ and the family of open subfunctors $\Hb_{u}$ is a covering of $\mathbf{Hilb}^{\mu}_{\PP^{n}}$. 
\end{proof}

\subsection{Representation of the Hilbert functor}

We are now going to prove that the Hilbert functor is representable.

\begin{definition}\label{hx0b}
Let $B$ be a family of $\mu$ monomials of degree $d$ in
$\mathbb{K}[x_{0},\ldots,x_{n}]$. Let $\Hb_{x_{0}}^{B}$ be the subfunctor of
$\Hb_{x_{0}}$ which associates to $X$ in $\mathcal{C}$ the set
$\Hb_{x_{0}}^{B}(X)$ of  flat families  $Z \subset X \times \mathbb{P}^{n}$
of closed subschemes of  $Y$ parametrized by  $X$ with fibers having Hilbert
polynomial  $\mu$  such that $\pi_{*}(O_{Z_{x_{0}}})$ is a locally free sheaf
of rank $\mu$ of X with basis $\underline{B}:=B/x_{0}^{d}$ considered as
elements of $H^{0}(Z,O_{Z_{x_{0}}})$. 
\end{definition} 

\begin{lemma}\label{lemme}
Let $d \geq \mu$ be an integer. Let $\mathcal{B}_{d}$ be the set of families $B$ of $\mu$ monomials of degree $d$ in $\mathbb{K}[x_{0},\ldots,x_{n}]$ such that the affinization $\underline{B}$ is connected to one in $\mathbb{K}[x_{1},\ldots,x_{n}]$. Then, the family of contravariant functors $(\Hb_{x_{0}}^{B})_{B\in \mathcal{B}_{d}}$ consists of open covering of representable subfunctors of $\Hb_{x_{0}}$.

\end{lemma}
\begin{proof}
First, let us prove that $\Hb_{x_{0}}^{B}$ is an open subfunctor. By proposition \ref{functor}, we can reduce to the case of affine schemes. Let $A$ and $A'$ be noetherian $\mathbb{K}$-algebras. Let $f$ be any morphism of $\mathbb{K}$-algebras from $A$ to $A'$ and $\phi$ its corresponding morphism from $\mathbf{Spec}(A')$ to $\mathbf{Spec}(A)$.\\
 Let $Z$ be an element of $\mathbf{Hilb}^{\mu}_{\mathbb{P}^{n}}(X) $ and $I$
 be its associated saturated homogeneous ideal of $A[x_{0},\ldots,x_{n}]$. Let
 $Z'$ be the element of $\mathbf{Hilb}^{\mu}_{\PP^{n}}(X')$ given by $(\phi
 \times  Id_{\mathbb{P}^{n}})^{*}(Z)$. Let $I'$ be the homogeneous ideal of
 $A'[x_{0},\ldots,x_{n}]$ associated to the quotient algebra
 $(A[x_{0},\ldots,x_{n}]/I) \otimes_{A} A'$.\\ 
Then, $Z'_{x_{0}} = (\phi \times Id_{\mathbb{P}^{n}})^{*}(Z_{x_{0}})$ is
equal to $\mathbf{Spec}(A'[x_{1},\ldots,x_{n}]/\underline{I}')$ (where
$\underline{ I }'$ denote the affinization of $I'$). 
Thus, we need to prove that for all noetherian $\mathbb{K}$-algebras
$A'$ and for all morphisms of $\mathbb{K}$-algebras $f$ from $A$ to $A'$,
the sheaf of modules of $\mathbf{Spec}(A')$ associated to the $A'$-module
$A'[x_{1},\ldots,x_{n}]/\underline{I}'$ is locally free of rank $\mu$ with basis
$\underline{B}$ if and only if the morphism  
$$
 \phi: \mathbf{Spec}(A')  \rightarrow \mathbf{Spec}(A) 
$$
factors through an open subscheme $\Gamma_{B}$ of $\mathbf{Spec}(A)$.\\
From proposition \ref{hu}, one has that $\Gamma_{B}$ exists and is equal to the open subscheme of $\Omega_{x_{0}}$ associated to the open subset on which the sheaf of module
given by the $A$-module $A[x_{1},\ldots,x_{n}]/\underline{I}$ is locally free of
rank $\mu$ with basis $\underline{B}$. Thus, $\Hb_{u}^{B}$ is an open subfunctor of $\Hb_{x_{0}}$.\\ 
To prove that the family $(\Hb_{u}^{B})_{B \in \mathcal{B}_{d}}$ is a
covering, we need to prove that the family $(\Gamma_{B})_{B \in
  \mathcal{B}_{d}}$ is a covering of $\mathbf{Spec}(A)$. This is a
straightforward consequence of the fact that any zero dimensional $k$-algebra
$k[x_{1},\ldots,x_{n}]/J$ (where $k$ is a field and $J$ an ideal of
$k[x_{1},\ldots,x_{n}]$) has a basis of monomials connected to one (take for
instance the complement of the initial ideal of $J$ for a monomial
ordering).\\

Finally we need to prove that $\Hb_{x_{0}}^{B}$ is representable. By proposition \ref{affine}, we can reduce to affine schemes.\\
Let $A$ be a noetherian $\mathbb{K}$-algebra and $X = \mathbf{Spec}(A)$. Recall that $\Hb_{x_{0}}^{B}(X)$ is the set of
saturated homogeneous ideals $I$ of $A[x_{0},\ldots,x_{n}]$ such that for all $d \geq \mu$ $(A[x_{0},\ldots,x_{n}] / I)_{d}$ is a flat $A$ module, for every prime ideal $p \subset A$, the Hilbert polynomial of the $k(p)$-graded algebra $(A[x_{0},\ldots,x_{n}]/I) \otimes_{A} k(p)$ is equal to $\mu$ and  $A_{p}[x_{1},\ldots,x_{n}] / \underline{I}_{p}$ is a free $A_{p}$-module of basis $\underline{B}$.\\

Let $F$ be the contravariant functor from the category of affine schemes to the category of Sets which associates to $X$ the set $F_{B}(X)$ of ideals $J$ of $A[x_{1},\ldots,x_{n}]$ such that  $A[x_{1},\ldots,x_{n}] /J$ is a free $A$ module of basis $\underline{B}$.\\

Let $\psi$ be the morphism of functors from $\Hb_{x_{0}}^{B}$ to $F$ given by
\begin{eqnarray*}
\psi: \Hb_{x_{0}}^{B}(X) &\longrightarrow& F_{B}(X)\\
I &\longmapsto& \underline{I}
\end{eqnarray*}
The map $\psi$ is a bijection whose inverse consists of the
homogenization. Then the functors $F_{B}$ and $\Hb_{x_{0}}^{B}$ are
isomorphic.

By proposition \ref{H}, $F_{B}$ is represented by
$\mathbf{Spec}(\mathbb{K}[(z_{\alpha,\beta})_{\alpha \in \delta B, \beta \in
    B}]/\mathcal{R})$, where $\mathcal{R}$ is the ideal generated by the
commutation relations \eqref{eq:commute}. Thus $\Hb_{x_{0}}^{B}$ is
representable in the category of affine schemes. By proposition \ref{affine},
$\Hb_{x_{0}}^{B}$ is representable in $\mathcal{C}$.
\end{proof} 

\begin{theorem}
 The contravariant functor $\mathbf{Hilb}^{\mu}_{\PP^{n}}$ from $\mathcal{C}$ to the category of Sets is representable.
\end{theorem}

\begin{proof}
From lemma \ref{lemme} and proposition \ref{represente}, $\Hb_{x_{0}}$ is a representable functor. More generally, by a change of variables in $\mathbb{K}[x_{0},\ldots,x_{n}]$, $\Hb_{u}$ is a representable functor for all $u \in \mathbb{K}[x_{0},\ldots,x_{n}]_{1}$.\\
Thus, from propositions \ref{hu} and \ref{represente}, $\mathbf{Hilb}^{\mu}_{\PP^{n}}$ is a representable contravariant functor.

\end{proof}

\begin{remark}
 Note that the $\mathbb{K}$-rational points of $\mathbf{Hilb}^{\mu}(\PP^{n})$ are by definition in bijection with the set of homogeneous saturated ideal $I$ of $\mathbb{K}[x_{0},\ldots,x_{n}]$ such that the quotient algebra $S/I$ has Hilbert polynomial equal to $\mu$. 
\end{remark}

\section{Global equations of the Hilbert Scheme}
Let $A$ be $\mathbb{K}$-algebra and a noetherian local ring of maximal ideal $m$ and residue field $k:= A/m$. Recall that if $a \notin m$ then $a$ is invertible in $A$.  Let $X:= \mathbf{Spec}(A)$ be the affine scheme associated to $A$ and $\mu$ be an integer. Let $T:=A[x_{0},\ldots,x_{n}]$ be the polynomial over $A$ in $n+1$ variables and $V:=A[x_{1},\ldots,x_{n}]$ the polynomial ring over $A$ in $n$ variables.

\subsection{Gotzmann's persistence and regularity theorems}\label{grassrepr}

Recall that from  corollary \ref{laremark}, $\mathbf{Hilb}^{\mu}_{\mathbb{P}^{n}}(X) $ is equal to set of homogeneous ideals $I$ of $T$ such that $T_{d}/I_{d}$ is a free $A$-module of rank $\mu$ for all $d \geq \mu$.\\
Finally, recall the definition of the Grasmmannian functor (see eg. \cite{sernesi}[Chap.4.3.3, p.209]): 
\begin{definition}
Let $V$ be a $\mathbb{K}$-vector space of finite dimension $N$ and $n\leq N$
an integer. Let $X$ be a noetherian scheme over $\mathbb{K}$. The $n$
Grassmannian functor of $V$ is the contravariant functor from the category
$\mathcal{C}$ to the category of Sets which associates to $X$ the set
$\mathbf{Gr}^{n}_{V}(X)$ of locally free sheaves  $\epsilon$ of rank  $n$
such that $\epsilon$ is a quotient of
$V^{*}\otimes_{\mathbb{K}}\mathcal{O}_{X}$ on X. 
\end{definition}
The following theorems come from \cite {Gotzmann}:
\begin{theorem}[Persistence theorem]\label{persis}
 Let $d, \mu$ be integers such that $d\geq \mu$. Let $B$ be any noetherian ring and $F= B[x_{0},\ldots,x_{n}]$. Let $I$ be an homogeneous ideal of $F$ generated by $I_{d}$ and let $M=F/I$. If $M_{i}$ is a flat $B$-module of rank $\mu$ for $i=d,d+1$, then $M_{i}$ is so for all $i \geq d$.
\end{theorem}

\begin{theorem}[Gotzmann's regularity theorem]\label{regul}
Let $I$ be an homogeneous ideal of $S$ with Hilbert polynomial $\mu$. Then $I$ is $\mu$ regular:
$$
\text{H}^{i}(\mathbb{P}^{n},\widetilde{I}(\mu-i)) = 0 
$$
for $i > 0$, where $\widetilde{I}$ is the quasi-coherent sheaf associated to $I$.
\end{theorem}
For definition of $H^{i}(\mathbb{P}^{n},-)$, see \cite{hartshorne}[Chap.III, \S 8].\\

Using Persistence theorem \ref{persis} and corollary \ref{regularite}, we deduce the following propositions:

\begin{proposition}\label{prop:1.5}
 Given an integer $d$ such that $d \geq \mu$, $\mathbf{Hilb}^{\mu}_{\mathbb{P}^{n}}(X) $ is in bijection with the subset $W$ of
 $\Gr^{\mu}_{S_{d}^{*}}(X)\times\Gr^{\mu}_{S_{d+1}^{*}}(X)$
 defined by 
$$W=\{ (T_{d}/I_{d},T_{d+1}/I_{d+1}) \in
 \Gr^{\mu}_{S_{d}^{*}}(X)\times\Gr^{\mu}_{S_{d+1}^{*}}(X)\ |\ 
 T_{1}.I_{d} = I_{d+1}\}.
$$ 
\end{proposition}

\begin{proposition}[\cite{MR1735271}]\label{prop:2.14}
 Given an integer $d$ such that $d \geq \mu$, $\mathbf{Hilb}^{\mu}_{\mathbb{P}^{n}}(X) $
 is in bijection with the subset $G$ of
 $\Gr^{\mu}_{S_{d}^{*}}(X)$ given by 
$$
G=\{ T_{d}/I_{d} \in \mathbf{Gr}^{\mu}_{S_{d}^{*}}(X)\ |\  T_{d+1}/(T_{1}.I_{d}) \text{ is a free } A\text{-module of rank } \mu \}.
$$
\end{proposition}

Actually, proposition \ref{prop:1.5} can be reformulated this way:

\begin{proposition}\label{prop:2.5}
 Given an integer $d$ such that $d \geq \mu$, $\mathbf{Hilb}^{\mu}_{\mathbb{P}^{n}}(X) $ is in bijection with the subset $W$ of
 $\Gr^{\mu}_{S_{d}^{*}}(X)\times\Gr^{\mu}_{S_{d+1}^{*}}(X)$
 defined by 
$$W=\{ (T_{d}/I_{d},T_{d+1}/I_{d+1}) \in
 \Gr^{\mu}_{S_{d}^{*}}(X)\times\Gr^{\mu}_{S_{d+1}^{*}}(X)\ |\ 
 T_{1}.I_{d} \subset I_{d+1}\}.
$$ 
\end{proposition}
\begin{proof}
 Thanks to proposition \ref{prop:1.5}, we only need to prove that if $(I_{d},I_{d+1})$, with $I_{d}.S_{1} \subset I_{d+1}$, belongs to $\Gr^{\mu}_{S_{d}^{*}}(X)\times\Gr^{\mu}_{S_{d+1}^{*}}(X)$ then $I_{d}.S_{1} = I_{d+1}$.\\
Consider $I_{d}(m)$ and $I_{d+1}(m)$ introduced in definition \ref{I(p)}. Then one has
$$
(A[x_{0},\ldots,x_{n}]_{j}/I_{j}) \otimes k \sim k[x_{0},\ldots,x_{n}]/I_{j}(m)
$$
for $j=d,d+1$. Thus, one has that $S_{1}.I_{d}(m) \subset I_{d+1}(m)$ and $\mathrm{dim}_{k} k[x_{0},\ldots,x_{n}]/I_{d}(m) = \mathrm{dim}_{k} k[x_{0},\ldots,x_{n}]/I_{d+1}(m) = \mu$ with $d \geq \mu$. Then, by minimal growth of an ideal in $k[x_{0},\ldots,x_{n}]$ (see \cite{MR1735271}[Cor.C.4, p.291]), we deduce $S_{1}.I_{d}(m)=I_{d+1}(m)$. Let $I(m)$ (resp. $\mathcal{M}$) be the ideal generated by $I_{d}(m)$ (resp. $S_{1}$) in $k[x_{0},\ldots,x_{n}]$. Thus, by the persistence theorem \ref{persis} and minimal growth \cite{MR1735271}[Cor.C.4, p.291], we get that 
$$
\mathrm{Sat}(I(m)) := \bigcup_{j \in \mathbb{N}} I(m):(\mathcal{M})^{j} =\overline{I_{d}(m)}
$$
with $\overline{I_{d}(m)}$ equal to $I(m)+(I_{d}(m):S_{1})+(I_{d}(m):S_{2})+\ldots+(I_{d}(m):S_{d-1})$, and that $\overline{I_{d}(m)}$ belongs to $\mathbf{Hilb}^{\mu}_{\mathbb{P}^{n}}(k)$. Then, as $\overline{I_{d}(m)}$ is saturated, we can assume by a change of coordinates in $\mathbb{K}[x_{0},\ldots,x_{n}]$ that $x_{0}$ is not a zero divisor of $\overline{I_{d}(m)}$. It implies that the multiplication by $x_{0}$:
$$
*x_{0}: k[x_{0},\ldots,x_{n}]/I_{d}(m) \longrightarrow k[x_{0},\ldots,x_{n}]/I_{d+1}(m)
$$
is injective (and by dimension is an isomorphism).\\
Thus one has the following diagram:
$$
\xymatrix{ 
   (A[x_{0},\ldots,x_{n}]_{d}/I_{d})\otimes k \ar[r]^{\sim} \ar[d]^{*x_{0}} &  k(p)[x_{0},\ldots,x_{n}]_{d}/I_{d}(m) \ar[d]^{*x_{0}} \\
   (A[x_{0},\ldots,x_{n}]_{d+1}/I_{d+1})\otimes k \ar[r]^{\sim} & k(p)[x_{0},\ldots,x_{n}]_{d+1}/I_{d+1}(m)
}
$$
in which the morphisms of multiplication by $x_{0}$ are isomorphisms. Thus, as $A[x_{0},\ldots,x_{n}]_{j}/I_{j}$ are free $A$-module for $j=d,d+1$, the morphism of multiplication by $x_{0}$:
$$\xymatrix{
A[x_{0},\ldots,x_{n}]_{d}/I_{d} \ar[r]^{*x_{0}} & A[x_{0},\ldots,x_{n}]_{d+1}/I_{d+1}
}
$$
is an isomorphism.\\
Then, we proceed by dehomogenization by $x_{0}$. One has that the natural inclusion
$$
A[x_{1},\ldots,x_{n}]_{\leq d}/\underline{I_{d}} \longrightarrow A[x_{1},\ldots,x_{n}]_{\leq d+1}/\underline{I_{d+1}}
$$
is an isomorphism. As $d \geq \mu = \mathrm{rank}_{A}
A[x_{1},\ldots,x_{n}]_{\leq d}/\underline{I_{d}}$, we can find a basis $B$ of
$A[x_{1},\ldots,x_{n}]_{\leq d}/\underline{I_{d}}$ of polynomials of degree
stricly less than $d$ and define operators of multiplications by the
variables $(x_{i})_{1\leq i \leq n}$ in $A[x_{1},\ldots,x_{n}]_{\leq
  d}/\underline{I_{d}}$. Finally, we conclude as in proof of corollary
\ref{regularite} that $S_{1}\,I_{d} = I_{d+1}$. 
\end{proof}

\begin{remark}

Note that the bijection introduced in proposition \ref{prop:2.5} is the following:
\begin{eqnarray*}
 \{ (I_{d},I_{d+1}) |  T_{1}.I_{d} \subset I_{d+1} \text{ and } T_{k}/I_{k} \text{ is free of rank} \ \mu \ \text{for } k=d,d+1 \} \rightarrow \mathbf{Hilb}^{\mu}_{\mathbb{P}^{n}}(X) 
\end{eqnarray*}
\begin{eqnarray*}
(I_{d},I_{d+1}) \mapsto \overline{I_{d}}
\end{eqnarray*}
where $\overline{I_{d}} = (I_{d})+(I_{d}:T_{1})+(I_{d}:T_{2})+\ldots+(I_{d}:T_{d-1})$.\\

The same way, the bijection introduced in proposition \ref{prop:2.14} is the following:
\begin{eqnarray}\label{bij2}
 \{ I_{d}\subset T_{d} | \  T_{d}/I_{d} \text{ is a free } A \text{-module of rank } \mu \} \rightarrow \mathbf{Hilb}^{\mu}_{\mathbb{P}^{n}}(X) 
\end{eqnarray}
\begin{eqnarray*}
(I_{d}) \mapsto \overline{I_{d}}
\end{eqnarray*}
\end{remark}

\begin{remark}\label{projection}

Note that the previous bijections induce the following commutative diagram:
\begin{equation}\label{eq:comm:diag1}
\xymatrix{ \mathbf{Hilb}^{\mu}_{\mathbb{P}^{n}}(X) \ar[rr]^{\psi} \ar[rd]^{\phi} && W \subset \Gr^{\mu}_{S_{d}^{*}}(X)\times\Gr^{\mu}_{S_{d+1}^{*}}(X)  \ar[ld]^{\pi} \\ & G \subset \mathbf{Gr}^{\mu}_{S_{d}^{*}}(X) }
\end{equation}
where $\phi$ and $\psi$ are bijections and $\pi$ is the natural projection on the first Grassmannian $\mathbf{Gr}^{\mu}_{S_{d}^{*}}(X)$. Thus, we deduce $\pi$ is also a bijection and $G^{'}$ is exactly the projection of $W^{'}$ on $\mathbf{Gr}^{\mu}_{S_{d}^{*}}(X)$.\\

\end{remark}

\subsection{Global description} \label{sec:2}
From the previous section, we can consider $\mathbf{Hilb}^{\mu}_{\mathbb{P}^{n}}(X) $ as a subset $W$ of the product of the two Grassmannians: $\Gr^{\mu}_{S_{d}^{*}}(X)\times\Gr^{\mu}_{S_{d+1}^{*}}(X)$ or as a subset $G$ of the single Grassmannian: $\Gr^{\mu}_{S_{d}^{*}}(X)$ for $d \geq \mu$. In this section we will prove that $\mathbf{Hilb}^{\mu}_{\mathbb{P}^{n}}(X) $ can actually be considered as an algebraic subvariety of this product of two Grassmannians or as an algebraic subvariety of this single Grassmannian. We will get the global equations of these subvarieties in the Pl\"ucker coordinates and will connect it to the border basis description of Section \ref{sec:bb}.\\
 \\
We consider the well known embedding of $\Gr^{\mu}_{S_{d}^{*}}(X)$ into the projective space $\mathbb{P}(\wedge^{\mu}T_{d}^{*})$ given as follow:
let $\Delta:=T_{d}/I_{d}$ be an element of $\Gr^{\mu}_{S_{d}^{*}}(X)$ and $(e_{1},\ldots,e_{\mu})$ be any basis of the free $A$-module $\Delta$. For any ordered family $(\xb^{\alpha_{1}},\ldots,\xb^{\alpha_{\mu}})$ of $\mu$ monomials of degree $d$ (for some monomial ordering $<$) write:
$$
\xb^{\alpha_{1}}\wedge \cdots \wedge \xb^{\alpha_{\mu}}= \Delta_{\alpha_{1},\ldots,\alpha_{\mu}} e_{1}\wedge \cdots\wedge e_{\mu}
$$
in $\wedge^{\mu} (T_{d}/I_{d}) $ which is free of rank 1 and has basis
$e_{1}\wedge \cdots\wedge e_{\mu}$ with $\Delta_{\alpha_{1},\ldots,\alpha_{\mu}}$
in $A$. Finally, let us associate to $\Delta \in \Gr^{\mu}_{S_{d}^{*}}(X)$ the point in $\mathbb{P}(\wedge^{\mu}T_{d}^{*})$ corresponding to the family $(\Delta_{\alpha_{1},\ldots,\alpha_{\mu}})_{\alpha_{1}<\ldots<\alpha_{\mu}}$ (note that this construction does not depend on the choice of the basis $(e_{1},\ldots,e_{\mu})$ of $\Delta$).\\
 The $(\Delta_{\alpha_{1},\ldots,\alpha_{\mu}})$ will be called the Pl\"ucker coordinates. They satisfy the well known Pl\"ucker relations.
\\

Let $\Delta$ be an element of $\mathbf{Gr}^{\mu}_{S_{d}^{*}}(X)$. Let
$(\delta_{1}, \ldots, \delta_{\mu})$ in $T_{d}^{*}$ be the dual basis of
$(e_{1},\ldots, e_{\mu})$ in the dual space $\Delta^{*}= \mathrm{Hom}_{A}(\Delta, A)$
which is also a free $A$-module of rank $\mu$. Then the Pl\"ucker coordinates
of $\Delta$ as an element of $\PP(\wedge^{\mu} T_{d}^{*})$ are given by:

$$
\Delta_{\beta_{1},\ldots,\beta_{\mu}} 
=
\left|
\begin{array}{ccc}
\delta_{1}(\xb^{\beta_{1}}) &\cdots & \delta_{1}(\xb^{\beta_{\mu}}) \\
\vdots & & \vdots  \\
\delta_{\mu}(\xb^{\beta_{1}}) &\cdots & \delta_{\mu}(\xb^{\beta_{\mu}}) \\
\end{array}
\right|
$$
for $\beta_{i} \in \mathbb{N}^{n+1}, \ |\beta_{i}|=d$ and $\beta_{1} < \cdots <\beta_{\mu}$.\\
More generally, consider the following determinant:
$$ 
\left|
\begin{array}{ccc}
\delta_{1}(p_{1}) &\cdots & \delta_{1}(p_{\mu}) \\
\vdots & & \vdots  \\
\delta_{\mu}(p_{1}) &\cdots & \delta_{\mu}(p_{\mu}) \\
\end{array}
\right|
$$ 
for $(p_{1},\ldots,p_{\mu})$ any family of polynomials in $S_{d}$ (not
necessarily monomials and not necessarily ordered). Using the multilinearity
properties of the determinant and the equality above, it is easy to prove
that this determinant can be written as a linear form in the Pl\"ucker
coordinates $(\Delta_{\beta_{1},\ldots,\beta_{\mu}})_{|\beta_{i}|=d,
  \beta_{1} < \cdots <\beta_{\mu}}$. We will denote it
$\Delta_{p_{1},\ldots,p_{\mu}}$ or $\Delta_{E}$ where
$E=(p_{1},\ldots,p_{\mu})$ is a family of polynomials in $T$.\\ 
For example, let $F=(\xb^{\alpha_{1}}, \ldots,\xb^{\alpha_{\mu}})$ be a
family of $\mu$ monomials in $T_{d}$ (not necessarily ordered). Then
$\Delta_{F} = \epsilon.\Delta_{F^{'}}$ where $F^{'}$ is the family of
monomials associated to $F$ ordered by the monomial ordering $<$, and
$\epsilon \in \{\pm 1\}$ is the signature of the permutation which transforms
$F$ into $F^{'}$.

From now on, if $\Delta \in \Gr^{\mu}_{S_{d}^{*}}(X)$, we will denote by
$\ker(\Delta)$ the $A$-submodule $I_{d}$ of $T_{d}$ such that $\Delta = T_{d}/I_{d}$.

We are going to describe the border relations with respect
to a basis $B \subset S_{d}$ of $\Delta:=T_{d}/I_{d}$ in terms of these Pl\"ucker
coordinates. 
\begin{lemma}\label{lemma:plucker:coord}
Let $\Delta:=T_{d}/I_{d}$ be an element of $\Gr^{\mu}_{S_{d}^{*}}(X)$. Let $B=(b_{1},\ldots,b_{\mu})$ be a family of polynomials of degree $d$. Then we have the following relation: 
$$
\Delta_{B}\,a-\sum_{i=1}^{\mu} \Delta_{B^{[b_{i}|a]}} 
 \,b_{i}=0 \ \mathrm{in} \ \Delta,  \mathrm{for}\ a \in T_{d}
$$ 
where $B^{[b_{i}|a]}=
(b_{1},\ldots,b_{i-1},a,b_{i+1},\ldots,b_{\mu})$. 
\end{lemma}
\begin{proof}
The previous relation is a straightforward consequence of basic properties of determinants. Consider the following matrix 
$$ 
M:=
\left[
\begin{array}{cccc}
\delta_{1}(a) & \delta_{1}(b_{1}) &\cdots & \delta_{1}(b_{\mu}) \\
\vdots & & & \vdots  \\
\delta_{\mu}(a) & \delta_{\mu}(b_{1}) &\cdots & \delta_{\mu}(b_{\mu}) \\
1 & 1 & \cdots & 1 \\
\end{array}
\right]
$$
and develop its determinant along the last row of $M$. Then one has the following relation
$$M
\left[
\begin{array}{c}
\Delta_{B} \\
\Delta_{B^{[b_{1}|a]}}  \\
\vdots   \\
\Delta_{B^{[b_{\mu}|a]}}\\
\end{array}
\right] = 
\left[
\begin{array}{c}
0 \\
0  \\
\vdots   \\
\mathrm{det}(M)\\
\end{array}
\right].
$$
From this relation, we conclude that $\Delta_{B}\,a-\sum_{i=1}^{\mu} \Delta_{B^{[b_{i}|a]}} 
 \,b_{i}=0$ in $\Delta$.
\end{proof}

\begin{theorem}
Let $d \geq \mu$ be an integer. $\mathbf{Hilb}^{\mu}_{\mathbb{P}^{n}}(X) $ is the projection on 
$\Gr^{\mu}_{S_{d}^{*}}(X)$ of the variety of $\Gr^{\mu}_{S_{d}^{*}}(X)\times \Gr^{\mu}_{S_{d+1}^{*}}(X)$ defined by the equations 
\begin{equation}\label{eq0:hilbert:scheme}
\Delta_{B}\, \Delta^{'}_{B',x_{k} a} - \sum_{b \in B} 
\Delta_{B^{[b|a]}}\,\Delta^{'}_{B',x_{k} b}=0,
\end{equation}
for all families $B$ (resp. $B'$) of $\mu$ (resp. $\mu-1$)
monomials of degree $d$ (resp. $d+1$), all monomial $a \in T_{d}$ and for every $k$ (where $B^{'},x_{k}a$ is the family $(b^{'}_{1},\ldots,b^{'}_{\mu-1},x_{k}a)$).
\end{theorem}
\begin{proof}
From remark \ref{projection} about the commutative diagram \eqref{eq:comm:diag1}, $G$ is equal to the projection of $W$ onto $\Gr^{\mu}_{S_{d}^{*}}(X)$.  Thus it is enough to prove that $W \subset \Gr^{\mu}_{S_{d}^{*}}(X)\times\Gr^{\mu}_{S_{d+1}^{*}}(X)$ is the subvariety defined by the equations \eqref{eq0:hilbert:scheme}. By proposition \ref{prop:2.5}, $W$ is defined by the single condition: $T_{1}.\ker \Delta \subset \ker \Delta^{'}$. Thus we need to prove \eqref{eq0:hilbert:scheme} is equivalent to $T_{1}.\ker \Delta \subset \ker \Delta^{'}$.\\
First of all, let $(\Delta,\Delta^{'})$ be an element of
$\Gr^{\mu}_{S_{d}^{*}}(X)\times \Gr^{\mu}_{S_{d+1}^{*}}(X)$ satisfying the equations
\eqref{eq0:hilbert:scheme}. We need to prove that $T_{1}\cdot \ker \Delta \subset \ker
\Delta^{'}$. Let $B$ be a basis of $\Delta$ (so that $\Delta_{B} \not\in m$ is
invertible), and let $p$ be an element of $\ker \Delta$. By linearity, equations \eqref{eq0:hilbert:scheme} imply
that $\Delta^{'}_{B',x_{k} p}=0$ for all $k=1,\ldots,n$ and all subset $B'$
of $\mu - 1$ monomials of degree $d+1$ (because $\Delta_{B^{[b|p]}} = 0$). Thus, by lemma \ref{lemma:plucker:coord}, $x_{k}\cdot p$ belongs to $\ker
\Delta^{'}$ for all $k=1,..,n$ and $S_{1}\cdot \ker \Delta \subset \ker
\Delta^{'}$.

Conversely, let $(\Delta,\Delta^{'})$ satisfy $T_{1}.\ker \Delta \subset \ker
\Delta^{'}$. Thus by proposition \ref{prop:2.5}, $(\Delta,\Delta^{'})$ is in $W^{'}$ and corresponds to a homogeneous saturated ideal $I$ in $\mathbf{Hilb}^{\mu}_{\mathbb{P}^{n}}(X) $ so that $I_{d}=\ker
\Delta$ and $I_{d+1}=\ker \Delta^{'}$. We are going to prove that the equations
\eqref{eq0:hilbert:scheme} are satisfied for $(\Delta,\Delta^{'})$.\\
This is a straightforward consequence to lemma \ref{lemma:plucker:coord}: for any family $(b_{i})$ of $\mu$ polynomials in $T_{d}$ and any polynomial $a \in T_{d}$, one has the following formula:
$$ 
a\, \Delta_{B} = \sum_{i} \Delta_{B^{[b_{i}|a]}} \, b_{i}
$$
in $\Delta$. As $T_{1}.\ker \Delta \subset \ker \Delta^{'}$, one also has
$$
a\,x_{k}\,\Delta_{B} = \sum_{i} \Delta_{B^{[b_{i}|a]}}\,b_{i}\,x_{k}
$$
in $\Delta^{'}$ for any $0 \leq k \leq n$.\\
Then by linearity, for any family $B'$ of $\mu-1$ monomials,
$$
\Delta_{B}\, \Delta^{'}_{B',x_{k} a} = \Delta^{'}_{B',x_{k}\,\Delta_{B}a} = \sum_{b \in B} 
\Delta_{B^{[b|a]}}\,\Delta^{'}_{B',x_{k} b}
$$
which are precisely equations \eqref{eq0:hilbert:scheme}.

\end{proof}

Hereafter, we describe the equations of $\mathbf{Hilb}^{\mu}_{\mathbb{P}^{n}}(X) $ as a variety of the single Grassmannian 
$\Gr^{\mu}_{S_{d}^{*}}(X)$ (i.e the equations of $G$ introduced in proposition \ref{prop:2.14}). This is also equal to the projection on $\Gr^{\mu}_{S_{d}^{*}}(X)$ of the variety defined by the equations
\eqref{eq0:hilbert:scheme} in $\Gr^{\mu}_{S_{d}^{*}}(X)\times \Gr^{\mu}_{S_{d+1}^{*}}(X)$. Let us introduce a generic linear form  
$\ub = u_{0} x_{0} + \cdots + u_{n} x_{n}$ where $u_{i}$ are parameters in $A$.

For any family $B=(b_{1},\ldots, b_{\mu}) $ in $ S_{d-1}$, we define 
$$
\Delta_{\ub \cdot B}= \det (\delta_{i}(\ub \cdot b_{j}))=
\sum_{\mathcal{I} \in \{0,\ldots,n\}^{\mu}} \ub^{(\mathcal{I})} \Delta_{\mathcal{I}\cdot B},
$$
where $\Delta_{\mathcal{I}\cdot B}=\Delta_{x_{\mathcal{I}_{1}} b_{1} ,\ldots,  x_{\mathcal{I}_{\mu}}b_{\mu}}$ 
and $(\mathcal{I})$ is the element of $\NN^{n+1}$ such that $\ub^{(\mathcal{I})}=u_{\mathcal{I}_{1}} \cdots u_{\mathcal{I}_{\mu}}$
for all $\mathcal{I}\in \{0,\ldots,n\}^{\mu}$. In this context, $\mathcal{I}_{i}$ will also be denoted $\mathcal{I}_{b_{i}}$ and more generaly $\mathcal{I}_{b}$ for $b \in B$.

For two families of monomials $B, B' \subset T_{d-1}$, we have 
$$
\Delta_{\ub \cdot B}\, \Delta_{\ub\cdot B'} = \sum_{\mathcal{K} \in \NN^{n+1}, |\mathcal{K}|=2\,\mu}
\ub^{\mathcal{K}}\, \sum_{\mathcal{I},\mathcal{J} \in \{0,\ldots,n\}^{\mu},\ (\mathcal{I})+(\mathcal{J})=\mathcal{K}} \Delta_{\mathcal{I}\cdot B}
\, \Delta_{\mathcal{J}\cdot B}.
$$

\begin{proposition}\label{prop2.5}
Let $d \geq \mu$ be an integer $\Delta \in \Gr^{\mu}_{S_{d}^{*}}(X)$ be an element of $\mathbf{Hilb}^{\mu}_{\mathbb{P}^{n}}(X) $. Then for all families $B$ of $\mu$ monomials of
degree $d-1$, for all $\mathcal{K}\in \NN^{n+1}$ with $|\mathcal{K}|=2\, \mu$, for all
$b, b'' \in B$ and all $1\leq i < j \leq n$, we have 
\begin{equation}\label{eq:hilbert:scheme}
\sum_{(\mathcal{I}) + (\mathcal{J}) =\mathcal{K}}\sum_{b'\in B} 
(\Delta_{\mathcal{I}\cdot B^{[x_{\mathcal{I}_{b'}}b'|x_{i}b]}}\, 
\Delta_{\mathcal{J}\cdot B^{[x_{\mathcal{J}_{b''}} b''|x_{j} b']}}
- 
\Delta_{\mathcal{I}\cdot B^{[x_{\mathcal{I}_{b'}}b'|x_{j}b]}}\,
\Delta_{\mathcal{J}\cdot B^{[x_{\mathcal{J}_{b''}}b''|x_{i}b']}}) =0.
\end{equation}
\end{proposition}
\begin{proof} 
Note that the relations \eqref{eq:hilbert:scheme} are obtained (using previous notations) as the
coefficients in $\ub$ of the relations:
\begin{equation}\label{eq:hilbert:scheme:u}
\sum_{b'\in B} 
(\Delta_{\ub \cdot B^{\,[\ub b'|x_{i}b]}}\,
\Delta_{\ub \cdot B^{\,[\ub b''|x_{j}b']}} 
-
\Delta_{\ub \cdot B^{\,[\ub b'|x_{j}b ]}}\,
\Delta_{\ub \cdot B^{\,[\ub b''|x_{i}b']}})=0.
\end{equation}
By proposition \ref{generic}, proving  \eqref{eq:hilbert:scheme}  is thus equivalent to
proving \eqref{eq:hilbert:scheme:u} for generic values of $\ub$ in $S_{1}=\kk[x_{0},\ldots,x_{n}]$.

Let $I$ be the homogeneous ideal in $\mathbf{Hilb}^{\mu}_{\mathbb{P}^{n}}(X)
$ associated to $\Delta \in \Gr^{\mu}_{S_{d}^{*}}(X)$. Let $\Delta' =
T_{d+1}/I_{d+1}$. As $d \geq \mu$, we know by propositions \ref{prop:2.14} and
\ref{prop:2.5} that  $\Delta$ and $\Delta'$ are free $A$ modules of rank
$\mu$. Let $F:=(e_{1},\ldots,e_{\mu})$ be a basis of $\Delta$. Let $P_{1} \in
A[\ub]$ be the polynomial in $\ub$ given by: 
$$
P_{1}(\ub) := \Delta'_{\ub \cdot F}.
$$
As $I$ is saturated, one can find $\mathbf{v} \in T_{1}$ such that
$\mathbf{v}$ is not a zero divisor of $I$ and thus $P_{1}(\mathbf{v}) \notin
m$. Thus the polynonial $\overline{P_{1}}$ in $T \otimes (A/m) =
k[x_{0},\ldots, x_{n}]$ is not equal to zero. Consequently,
$\Delta'_{\ub.F} \notin m$ for generic values of $\ub$ in $S_{1}$. Let us
choose a generic $\ub$ such that $\Delta'_{\ub.F} \notin m$. Finally by a change of variables in $\mathbb{K}[x_{0},\ldots,x_{n}]$, we can assume that $u = x_{0}$.\\
Then $F:=(e_{1},\ldots, e_{\mu})$ and $x_{0}\cdot
F:=(x_{0}\,e_{1},\ldots, x_{0}\,e_{\mu})$ are basis of respectively $\Delta$
and $\Delta'$. Thus, for any family $B$ of $\mu$ monomials of  degree $d-1$
one has: 
$$
{\Delta_{x_{0}\cdot B}\over \Delta_{F}} = {\Delta'_{x_{0}^{2}\cdot
    B}\over\Delta'_{x_{0}\cdot F}}
$$ 
because for any homogeneous polynomial $a$  of degree $d$  the following decomposition in $\Delta$:
$$
a = \sum_{i} z_{i}\,e_{i}
$$
($z_{i} \in A$) 
 induces this decomposition in $\Delta '$ (since $x_{0} \text{Ker} \Delta \subset \text{ker} \Delta '$):
$$
x_{0}\,a = \sum_{i} z_{i}\,x_{0}\, e_{i}
$$
For, any family $B$ of $\mu$ monomials of degree $d-1$ and any $b',b,b'' \in B$, we have
$$
\sum_{b'\in B} 
\Delta_{x_{0} \cdot B^{\,[x_{0}b'|x_{i}b]}}\,\Delta_{x_{0} \cdot B^{\,[x_{0}b''|x_{j}b']}}  = {\Delta_{F}\over\Delta'_{x_{0}F}} \sum_{b'\in B} 
\Delta_{x_{0} \cdot B^{\,[x_{0}b'|x_{i}b]}}\,\Delta'_{x_{0}^{2} \cdot B^{\,[x_{0}^{2} b''|x_{0}x_{j}b']}}
$$
But by lemma \ref{lemma:plucker:coord} we have
$$
\sum_{b'\in B} x_{0}\,x_{j}b^{'}\,\Delta_{x_{0}\cdot B^{ \,[x_{0}b'|x_{i}b]}}
= x_{j}\,x_{i}b\,\Delta_{x_{0}\cdot B}.
$$
in $\Delta'$. \\
Finally, by linearity we get
$$
\sum_{b'\in B} 
\Delta_{x_{0} \cdot B^{\,[x_{0} b'|x_{i}b]}}\,\Delta_{x_{0} \cdot
  B^{\,[x_{0}b''|x_{j}b']}}  = {\Delta_{F}\over\Delta'_{x_{0}\cdot
    F}}\,\Delta_{x_{0}\cdot B}\,\Delta'_{x_{0}^{2} \cdot B^{\,[x_{0}^{2}b''|x_{i}x_{j}b]}}
$$
As this expression is symmetric in $i$ and $j$, we get
$$
\sum_{b'\in B} 
\Delta_{\ub \cdot B^{\,[\ub b'|x_{i}b]}}\,
\Delta_{\ub \cdot B^{\,[\ub b''|x_{j}b']}} =  \sum_{b'\in B} \Delta_{\ub \cdot B^{\,[\ub b'|x_{j}b ]}}\,
\Delta_{\ub \cdot B^{\,[\ub b''|x_{i}b']}},
$$
which proves the relations \eqref{eq:hilbert:scheme:u}.
\end{proof}
\begin{remark}\label{rmk}
\rm
Note that if $B$ is a basis of $\Delta$ and $\ub=x_{0}$ is not a zero divisor of $I$, then $\underline{B}$ is a basis of the quotient:
$$
\mathcal{A} := A[x_{1},\ldots,x_{n}]/\underline{I}
$$
and equations \eqref{eq:hilbert:scheme:u} are equivalent to the commutation relations between the operators of multiplication by the variable $(x_{i})$ in the quotient algebra $\mathcal{A}$.
\end{remark}

\begin{proposition}\label{prop2}
Let $d \geq \mu$ be an integer and $\Delta \in \Gr^{\mu}_{S_{d}^{*}}(X)$ be an element of
$\mathbf{Hilb}^{\mu}_{\mathbb{P}^{n}}(X) $. Then for all families $B$ of $\mu$
monomials of degree $d-1$, for all $\mathcal{K}\in \NN^{n+1}$ with $|\mathcal{K}|=2\, \mu$, for all monomial $a\in S_{d-1}$,
for all $b\in B$ and for all $k =0,\ldots ,n$, we have
\begin{equation}\label{eq2:hilbert:scheme}
\sum_{(\mathcal{I}) + (\mathcal{J}) =\mathcal{K}} \left(\Delta_{\mathcal{I}\cdot
  B^{[x_{\mathcal{I}_{b}}b|x_{k}a]}}\,\Delta_{\mathcal{J}\cdot B^{\ }} - \sum_{b' \in B} 
\Delta_{\mathcal{I}\cdot B^{[x_{\mathcal{I}_{b}}b|x_{k}b']}}\,\Delta_{\mathcal{J}\cdot B^{[x_{\mathcal{J}_{b'}}b'|x_{\mathcal{J}_{b'}}a]}}\right)=0.
\end{equation}
\end{proposition}
\begin{proof} 
Here also the relations \eqref{eq2:hilbert:scheme} are obtained as the
coefficients in $\ub$ of the relations
\begin{equation}\label{eq2:hilbert:scheme:u}
\Delta_{\ub\cdot
  B^{[\ub b|x_{k}a]}}\,\Delta_{\ub \cdot B} - \sum_{b' \in B} 
\Delta_{\ub \cdot B^{[\ub b|x_{k}b']}}\,\Delta_{\ub \cdot B^{[\ub b'|\ub a]}}=0.
\end{equation}
By proposition \ref{generic}, proving \eqref{eq2:hilbert:scheme}  is equivalent to
proving \eqref{eq2:hilbert:scheme:u} for generic values of $\ub \in S_{1}$.

Let $I$ be the homogeneous ideal in $\mathbf{Hilb}^{\mu}_{\mathbb{P}^{n}}(X)
$ associated to $\Delta \in \Gr^{\mu}_{S_{d}^{*}}(X)$. Let $\Delta' =
T_{d+1}/I_{d+1}$. As $d \ge \mu$, we know by propositions \ref{prop:2.14} and
\ref{prop:2.5} that  $\Delta$ and $\Delta'$ are free $A$ modules of rank
$\mu$. Let $F:=(e_{1},\ldots,e_{\mu})$ be a basis of $\Delta''$.\\
As we did in the proof of proposition \ref{prop2.5}, we can assume $\ub =
x_{0}$ and $F:=(e_{1},\ldots,e_{\mu})$ and
$x_{0}.F:=(x_{0}\,e_{1},\ldots,x_{0}\,e_{\mu})$ are basis of respectively
$\Delta$ and $\Delta'$. Then, for any family $B$ of $\mu$ monomials of
degree $d-1$ one has: 
$$
{\Delta_{x_{0} \cdot B}\over \Delta_{F}} = {\Delta'_{x_{0}^{2}\cdot
    B}\over\Delta'_{x_{0} \cdot F}}
$$
Thus for any family $B$ of $\mu$ monomials of degree $d-1$, any monomials $a$ of degree $d-1$ and any $k \in \mathbb{N}$:
$$
\Delta_{x_{0} \cdot B^{[x_{0} \cdot b|x_{k}b']}} = {\Delta_{F} \over
  \Delta'_{x_{0}\cdot F}}\,\Delta'_{x_{0}^{2} \cdot B^{[x_{0}^{2} b|x_{k}x_{0}b']}}.
$$
By lemma \ref{lemma:plucker:coord} we also have:
$$ 
\sum_{b' \in B} \Delta_{x_{0} \cdot B^{[x_{0} b'|x_{0} a]}}\,x_{0}\,b'\,x_{k}
= \Delta_{x_{0}\cdot B}\,x_{k}\, a
$$
in $\Delta'$.\\
Finally, by linearity we have:
$$
\sum_{b' \in B} \Delta_{x_{0} \cdot B^{\,[x_{0} b|x_{k}b']}}\,\Delta_{x_{0}
  \cdot B^{\,[x_{0} b'|x_{0} a]}} = {\Delta_{F} \over \Delta'_{x_{0}\cdot
    F}}\,\Delta_{x_{0}\,B}\,\Delta'_{x_{0}^{2}\cdot B^{\,[x_{0}^{2} b | x_{k}x_{0}a]}} = \Delta_{x_{0}\cdot B^{\,[x_{0} b|x_{k}a]}}\,\Delta_{x_{0} \cdot B}
$$
\end{proof} 
\begin{remark}\label{rmk2}
\rm
Note that if $B$ is a basis of $\Delta$ and $\ub = x_{0}$ is not a zero divisor of $I$, then $\underline{B}$ is a basis of the affinization of $T/I$:
$$
\mathcal{A} := A[x_{1},\ldots,x_{n}]/\underline{I}
$$
and equations \eqref{eq2:hilbert:scheme:u} are equivalent to the fact that the decomposition of $x_{k}a$ on $\underline{B}$ in $\mathcal{A}$ is obtained by computing the decomposition of $a$ on $\underline{B}$ and applying to it the operator of multiplication by $x_{k}$.
\end{remark}

\begin{example}
Here is an example of equations that are obtained from propositions
\ref{prop2.5} and \ref{prop2} in the case $n=2$, $S=\mathbb{K}[x,y,z]$ and
$\mu = 2$.\\ 
We will give the equations given by \eqref{eq:hilbert:scheme:u} and
\eqref{eq2:hilbert:scheme:u} in the case $\ub = x$. There are three families
of two monomials of degree $\mu-1=1$: 
$$
B_{1}= (x,y), B_{2}=(x,z), B_{3}=(y,z).
$$
Thus,
$$
x\cdot B_{1}= (x^{2},xy), x\cdot B_{2}=(x^{2},xz), x\cdot B_{3}=(xy,xz).
$$
The variables of multiplications are $y$ and $z$. 
Let us focus on the case $B=B_{1}$. 
By homogeneisation of the equations of example \ref{ex:mu=2}:
$$
\Delta_{x\cdot B_{1}}\, M_y = \left( \begin{array}{cc}
     0 & \Delta_{xy,y^{2}} \\
     \Delta_{x^{2},xy} & \Delta_{x^{2},y^{2}} \\
   \end{array} \right),
\ 
\Delta_{x\cdot B_{1}}\, M_z = \left( \begin{array}{cc}
     \Delta_{xy,xz} & \Delta_{xy,yz} \\
     \Delta_{x^{2},xz} & \Delta_{x^{2},yz} \\
   \end{array} \right).
$$
we obtain a first set of equations of type \eqref{eq:hilbert:scheme:u} given
by the commutation property:
$$ 
\left\{\begin{array}{l}
\Delta_{y^{2},xy}\Delta_{x^{2},xz}-\Delta_{xy,yz}\Delta_{x^{2},xy}=0,\\
\Delta_{y^{2},xy}\Delta_{x^{2},yz}-\Delta_{xy,xz}\Delta_{xy,y^{2}}-\Delta_{xy,yz}\Delta_{x^{2},y^{2}}=0,
\end{array}
\right.
$$
Notice that these equations are not enough to define the Hilbert scheme,
since it is irredicble of dimension $4$ but these equations vanish when
$\Delta_{y^{2},xy}=0, \Delta_{x^{2},xy}=0,\Delta_{x^{2},y^{2}}=0$. 
The second set of equations of type \eqref{eq2:hilbert:scheme:u} in the case
$\ub=x$ and $B=B_{1}$, is given by the decomposition in $x \cdot B_{1}$ of the
monomials $v \,a$ and by the operator of multiplication by $v$ applied
to $x\,a$ with $a=x,y,z$ and $v = y,z$.
For instance, for $v=y$ and $a=z$ we have:
$$
\Delta_{x\cdot B_{1}}^{2}\, yz \equiv \left( \begin{array}{cc}
     \Delta_{x^{2},xy}\Delta_{xy,yz} \\
     \Delta_{x^{2},xy}\Delta_{x^{2},yz}\\
   \end{array} \right)
$$
and 
$$
\Delta_{x\cdot B_{1}}xz \equiv \left( \begin{array}{cc}
     \Delta_{xy,xz} \\
     \Delta_{x^{2},xz}\\
   \end{array} \right).
$$
Then, we write
$$
\Delta_{x\cdot B_{1}}^{2}\,yz - \Delta_{x\cdot B_{1}}^{2}M_{y}(xz) \equiv 0
$$
and we get
$$ \tiny 
\left\{\begin{array}{l}
\Delta_{x^{2},xy}\Delta_{xy,yz}-\Delta_{x^{2},xz}\Delta_{y^{2},xy}=0,\\
\Delta_{x^{2},xy}\Delta_{x^{2},yz}-\Delta_{xy,xz}\Delta_{x^2,xy}-\Delta_{x^{2},xz}\Delta_{x^{2},y^2}=0.
\end{array}
\right.
$$
We do the same for $v=z$ and obtain:
$$ \tiny
\left\{\begin{array}{l}
\Delta_{x^{2},xy}\Delta_{z^2,xy}-\Delta_{xy,xz}\Delta_{xy,xz}-\Delta_{x^2,xz}\Delta_{xy,yz}=0,\\
\Delta_{x^{2},xy}\Delta_{x^{2},z^2}-\Delta_{xy,xz}\Delta_{x^2,xz}-\Delta_{x^{2},xz}\Delta_{x^{2},yz}=0.\\
\end{array}
\right.
$$
If we do the same for $a=x,y$ we do not get any new
equations. Consequently, the equations obtained from relations
\eqref{eq:hilbert:scheme:u} and \eqref{eq2:hilbert:scheme:u} in the case
$\ub=x$ and $B=B_{1}=(x,y)$ are: 
$$ \tiny
\left\{\begin{array}{l}
\Delta_{y^{2},xy}\Delta_{x^{2},xz}-\Delta_{xy,yz}\Delta_{xy,x^{2}}=0,\\
\Delta_{y^{2},xy}\Delta_{x^{2},yz}-\Delta_{xy,xz}\Delta_{xy,y^{2}}-\Delta_{xy,yz}\Delta_{x^{2},y^{2}}=0, \\
\Delta_{x^{2},xy}\Delta_{xy,xz}+\Delta_{x^{2},y^{2}}\Delta_{x^{2},xz}-\Delta_{x^{2},yz}\Delta_{x^{2},xy}=0,\\
\Delta_{x^{2},xy}\Delta_{xy,z^2}-\Delta_{xy,xz}^{2}-\Delta_{x^2,xz}\Delta_{xy,yz}=0,\\
\Delta_{x^{2},xy}\Delta_{x^{2},z^2}-\Delta_{xy,xz}\Delta_{x^2,xz}-\Delta_{x^{2},xz}\Delta_{x^{2},yz}=0.\\
\end{array}
\right.
$$

In the case $B=B_{2}$, we get the equations by permutation of $y$ and $z$.\\
Finally, in the case $B=B_{3}$, we get:
 $$ \tiny
\left\{\begin{array}{l}
\Delta_{y^{2},xz}\Delta_{yz,xz} + \Delta_{yz,xz}\Delta_{xy,yz}-\Delta_{yz,xz}\Delta_{y^{2},xz}-\Delta_{z^2,xy}\Delta_{xy,y^2}=0,\\
\Delta_{xy,y^2}\Delta_{yz,xz} + \Delta_{xy,yz}^{2}-\Delta_{xy,yz}\Delta_{y^{2},xz}-\Delta_{xy,z^2}\Delta_{xy,y^2}=0, \\
\Delta_{y^{2},xz}\Delta_{xy,z^2} + \Delta_{xz,yz}\Delta_{xy,z^2}-\Delta_{xz,yz}^{2}-\Delta_{xy,z^2}\Delta_{xy,yz}=0,\\
\Delta_{xy,y^2}\Delta_{z^2,xy} + \Delta_{xy,yz}\Delta_{xy,z^2}-\Delta_{xy,yz}\Delta_{yz,xz}-\Delta_{xy,z^2}\Delta_{xy,yz}=0.
\end{array}
\right.
$$
from relation \eqref{eq:hilbert:scheme:u}, and
$$  \tiny
\left\{\begin{array}{l}
\Delta_{xy,xz}^2-\Delta_{y^{2},xz}\Delta_{x^{2},xz}-\Delta_{yz,xz}\Delta_{xy,x^2}=0,\\
\Delta_{xy,y^2}\Delta_{x^{2},xz}+\Delta_{xy,yz}\Delta_{xy,x^2}=0,\\
\Delta_{yz,xz}\Delta_{x^{2},xz}+\Delta_{z^2,xz}\Delta_{xy,x^2}=0,\\
\Delta_{xy,xz}^2-\Delta_{xy,yz}\Delta_{x^{2},xz}-\Delta_{xy,z^2}\Delta_{xy,x^2}=0.
\end{array}
\right.
$$
from relation \eqref{eq2:hilbert:scheme:u}.\\
Finally, the equations that are obtained from \eqref{eq:hilbert:scheme:u} and \eqref{eq2:hilbert:scheme:u} in the case $\ub = x$ are:
$$ \tiny
\left\{\begin{array}{l}
\Delta_{y^{2},xy}\Delta_{x^{2},xz}-\Delta_{xy,yz}\Delta_{x^{2},xy}=0,\\
\Delta_{y^{2},xy}\Delta_{x^{2},yz}-\Delta_{xy,xz}\Delta_{y^{2},xy}-\Delta_{xy,yz}\Delta_{x^{2},y^{2}}=0, \\
\Delta_{x^{2},xy}\Delta_{xy,xz}+\Delta_{x^{2},y^{2}}\Delta_{x^{2},xz}-\Delta_{x^{2},yz}\Delta_{x^{2},xy}=0,\\
\Delta_{x^{2},xy}\Delta_{z^2,xy}-\Delta_{xy,xz}^{2}-\Delta_{x^2,xz}\Delta_{xy,yz}=0,\\
\Delta_{x^{2},xy}\Delta_{x^{2},z^2}-\Delta_{xy,xz}\Delta_{x^2,xz}-\Delta_{x^{2},xz}\Delta_{x^{2},yz}=0,\\

\Delta_{z^{2},xz}\Delta_{x^{2},xy}-\Delta_{xz,yz}\Delta_{x^{2},xz}=0,\\
\Delta_{z^{2},xz}\Delta_{x^{2},zy}-\Delta_{xy,xz}\Delta_{z^{2},xz}-\Delta_{zy,xz}\Delta_{x^{2},z^{2}}=0, \\
\Delta_{x^{2},xz}\Delta_{xy,xz}+\Delta_{x^{2},z^{2}}\Delta_{x^{2},xy}-\Delta_{x^{2},zy}\Delta_{x^{2},xz}=0,\\
\Delta_{x^{2},xz}\Delta_{y^2,xz}-\Delta_{xy,xz}^{2}-\Delta_{x^2,xy}\Delta_{xz,yz}=0,\\
\Delta_{x^{2},xz}\Delta_{x^{2},y^2}-\Delta_{xy,xz}\Delta_{x^2,xy}-\Delta_{x^{2},xy}\Delta_{x^{2},zy}=0,\\

\Delta_{y^{2},xz}\Delta_{yz,xz} + \Delta_{yz,xz}\Delta_{xy,yz}-\Delta_{yz,xz}\Delta_{y^{2},xz}-\Delta_{z^2,xy}\Delta_{xy,y^2}=0,\\
\Delta_{xy,y^2}\Delta_{yz,xz} + \Delta_{xy,yz}\Delta_{xy,yz}-\Delta_{xy,yz}\Delta_{y^{2},xz}-\Delta_{xy,z^2}\Delta_{xy,y^2}=0, \\
\Delta_{y^{2},xz}\Delta_{z^2,xy} + \Delta_{yz,xz}\Delta_{xy,z^2}-\Delta_{yz,xz}\Delta_{yz,xz}-\Delta_{z^2,xy}\Delta_{xy,yz}=0,\\
\Delta_{xy,y^2}\Delta_{z^2,xy} + \Delta_{xy,yz}\Delta_{xy,z^2}-\Delta_{xy,yz}\Delta_{yz,xz}-\Delta_{xy,z^2}\Delta_{xy,yz}=0,\\
\Delta_{xy,xz}^2-\Delta_{xz,y^{2}}\Delta_{x^{2},xz}-\Delta_{xz,yz}\Delta_{x^{2},xy}=0,\\
\Delta_{xz,yz}\Delta_{xz,x^{2}}+\Delta_{xz,z^2}\Delta_{xy,x^2}=0,\\
\Delta_{xy,xz}^2-\Delta_{xy,yz}\Delta_{x^{2},xz}-\Delta_{xy,z^2}\Delta_{xy,x^2}=0.
\end{array}
\right.
$$
A complete set of equations of $\mathbf{Hilb}^{2}_{\mathbb{P}^{2}}$ is obtained by permutation of $x$, $y$ and $z$ in
these equations. Notice that such quadratic equations have also been computed
by Gr\"obner basis techniques for $\mathbf{Hilb}^{2}_{\mathbb{P}^{2}}$ in
\cite[p. 3]{BS10}.
\end{example}

\begin{theorem}\label{equi}
Let $d \geq \mu$. An element $\Delta \in \Gr^{\mu}_{S_{d}^{*}}(X)$  is in $\mathbf{Hilb}^{\mu}_{\mathbb{P}^{n}}(X) $ iff it satisfies the relations 
\eqref{eq:hilbert:scheme} and \eqref{eq2:hilbert:scheme}.
\end{theorem} 
\begin{proof}

First, let us prove that there exists a family $B$ of $\mu$ monomials of
degree $d-1$ and a linear form $\ub$ such that $\ub \cdot B$ is a basis of
$\Delta$.  

Let $I_{d} = \ker \Delta \subset T_{d}$. Tensoring by the residue field $k$,
we can reduce to the case $A=k$ and thus $T=S=k[x_{0},\ldots,x_{n}]$. To
prove that there exists a family $B$ of $\mu$ monomials of degree $d-1$ and a
linear form $\ub$ such that $\ub\cdot B$ is a basis of $\Delta$, it is enough
to prove this result for $\Delta_{in}:=S_{d}/J_{d}$ with $J$ equal to the
initial ideal of $(I_{d})$ (for the degree reverse lexicographic ordering $<$
such that $x_{i}>x_{i+1}$, $i=0,\ldots,n-1$). From
\cite{MR1322960}[Thm. 15.20, p. 351], by a generic change of variables, we can
assume that $J$ is Borel-fix i.e if $x_{i}x^{\alpha} \in J$ then
$x_{j}x^{\alpha} \in J$ for all $j > i$.

Let us prove now that there exists a family $B$ of $\mu$ monomials of degree
$d-1$ such that $x_{0}\cdot B$ is a basis of $S_{d}/J_{d}$. It is enough to prove
$J_{d} + x_{0}\,S_{d-1} = S_{d}$. Consider $J'_{d} :=
(J_{d}+x_{0}\,S_{d-1})/x_{0}\,S_{d-1}$ as a subvector space of $S'_{d}:=
S_{d}/x_{0}\,S_{d-1}$ (which is isomorphic to $k[x_{1},\ldots,x_{n}]_{d}$). We
need to prove that $S'_{d} = J'_{d}$. Let $L \subset S_{d-1}$ be the
following set
$$
L:= \{x \in S_{d-1} | \ x_{0}x \in J_{d} \}= (J_{d}:x_{0}).
$$
One has the exact sequence
$$
\xymatrix{
0 \ar[r] & S_{d-1}/L \ar[r]^{*x_{0}} & S_{d}/J_{d} \ar[r] & S'_{d}/J'_{d} \ar[r] & 0.
}
$$ Assume that $\dim (S'_{d}/J'_{d}) > 0$, then $\dim(L)>s_{d-1} -
\mu$. Thus, as $d \geq \mu$, by \cite{Gotzmann}[(2.10), p.66], $\dim(S_{1}\cdot
L) > s_{d}-\mu$. As $J$ is Borel-fix, $S_{1} \cdot L \subset J_{d}$. Thus,
$\dim(J_{d}) \geq \mathrm{dim}(S_{1}\cdot L) > s_{d}-\mu$. By assumption,
this is impossible, thus $J'_{d} = S'_{d}$.  we deduce that there exists a
family $B$ of $\mu$ monomials of degree $d-1$ such that $x_{0}\cdot B$ is a
basis of $S_{d}/J_{d}$.

Let $B$ be a family of $\mu$ monomials of degree $d-1$ such that $x_{0}\cdot B$ is
a basis of $\Delta=T_{d}/I_{d}$. Let $\underline{I}_{d} \subset V_{\leq
  d}=A[x_{1},\ldots,x_{n}]_{\leq d}$ be the dehomogenization of $I_{d}$ (the
free $A$ submodule of rank $\mu$ of $V_{\leq d}$ defined by putting $x_{0}=1$
in $I_{d}$).  Let $\pi: T_{d}/I_{d} \rightarrow \langle B \rangle$ be the
natural isomorphism of $A$ modules between $T_{d}/I_{d}$ and $\langle B
\rangle$ (the free $A$-module of basis $B$). The dehomogenization is also an
isomorphism of $A$-modules between $V_{\leq d}/\underline{I}_{d}$ and
$T_{d}/I_{d}$. Thus $V_{\leq d}/\underline{I}_{d}$ is naturally isomorphic to
$\langle B \rangle $. Let $\psi$ be this isomorphism, we have the following
commutative diagram:
$$
\xymatrix{ V_{\leq d}/\underline{I}_{d} \ar[rr]^{\sim} \ar[rd]^{\psi} && T_{d}/I_{d}  \ar[ld]^{\pi} \\ & \langle B \rangle }
$$

We introduce the linear operators $(M_{i})_{i=1,\ldots,n}$ operating on
$\langle B \rangle$. From remarks \ref{rmk} and \ref{rmk2}, relations \eqref{eq:hilbert:scheme:u} and
\eqref{eq2:hilbert:scheme:u} for $\ub = x_{0}$ give us that the operators
$(M_{i})_{i=1,\ldots,n}$ commute and that, for every $\alpha \in \mathbb{N}^{n+1}$
 with $|\alpha| = d-1$ and every $i=1,\ldots,n$, $\pi ( \xb^{\alpha}\,x_{i} ) =
M_{i}( \pi (\xb^{\alpha}\, x_{0} ) )$ (i.e $\psi ( \xbb^{\alpha}\,x_{i} ) =
M_{i}( \psi (\xbb^{\alpha} ) )$). Note that, by definition in section Notations \ref{notation}, for all $\alpha \in \NN^{n+1}$ such that $|\alpha| = d$ we have
$$
\pi(\xb^{\alpha}) = \psi(\xbb^{\alpha}).
$$

We define $\sigma$, an application from $V = A[x_{1},\ldots,x_{n}]$ to $\langle B \rangle$, as follows: $\forall p \in
V, \sigma(p) = p(M)(\pi(x_{0}^{d}))$ (or $p(M)(\psi(1))$) where $\xbb^{\alpha}(M) =
M_{1}^{\alpha_{1}}\cdots M_{n}^{\alpha_{n}}$. As the operators $(M_{i})_{i=1,\ldots,n}$ commute, $p(M)$
is well defined and $\Jc = \ker \sigma$ is an ideal of $V$. Let us prove by
induction on the degree $k$ of $p$ that $\sigma(p) = \psi(p)$ for polynomials $p$
in $V_{\leq d}$. For $k=0$, $\sigma(1) = \psi(1)$ by definition. From $k$ to $k+1$, we assume that $p$ is a monomial of degree $k+1$ i.e $p$ is of the form:
$$
p = x_{i}\,\xbb^{\alpha}
$$ 
with $i \in 1 \ldots n$ and $\alpha \in \NN^{n+1}$ such that $|\alpha| = d$ and degree of $\xbb^{\alpha}$ is equal $k$.
Then, $\sigma(x_{i}\,\xbb^{\alpha}) = M_{i}(\sigma(\xbb^{\alpha}))$. 
By induction on $k$, $\sigma(\xbb^{\alpha}) = \psi(\xbb^{\alpha})$. Thus, we have $\sigma(x_{i}\,\xbb^{\alpha}) = M_{i}(\psi(\xbb^{\alpha})) = M_{i}(\pi(\xb^{\alpha}))$. From the end of the previous paragraph, this is equal to
$\psi(x^{\alpha} \, x_{i}) = \psi(p)$.

Then it follows that $\sigma$ is surjective because for all
$\xb^{\alpha} \in B, \sigma(\xbb^{\alpha})=\psi (\xbb^{\alpha})=\pi(\xb^{\alpha}) = \xb^{\alpha}$. Thus we get that $V/\Jc$ is a free $A$-module of rank
$\mu=|B|$. Moreover, since $\sigma$ and $\psi$ coincide on $V_{\le d}$, $\Jc_{\le d} =
\underline{I_{d}}$. Let $J$ be the homogenization of $\Jc$, then by proposition \ref{prop:2.5} $J$ belongs to $\mathbf{Hilb}^{\mu}_{\mathbb{P}^{n}}(X) $ and $J_{d} = I_{d}$ (because $\Jc_{\le d} = \underline{I_{d}}$).

\end{proof}

Now, using theorem \ref{equi}, we can finally give equations for the Hilbert scheme $\mathrm{Hilb}^{\mu}(\PP^{n})$.\\ 

By definition $\mathrm{Hilb}^{\mu}(\PP^{n}) $ represents the Hilbert functor
$\mathbf{Hilb}^{\mu}_{\PP^{n}}$. We can reduce to the case of affine schemes
$X = \mathbf{Spec}(A)$, with $A$ a noetherian $\mathbb{K}$-algebra. In the following, we will say that an $A$ module $M$ is locally free of rank $r$ if the quasi coherent sheaf of modules denoted $\widetilde{M}$ on
$X = \mathbf{Spec}(A)$ is locally free of rank $r$.

Recall that the Hilbert functor associates to $X$ in the category
$\mathcal{C}$ of noetherian schemes over $\mathbb{K}$, the set
$\mathbf{Hilb}^{\mu}_{\mathbb{P}^{n}}(X) $ of saturated homogeneous ideals
$I$ of $A[x_{0},\ldots,x_{n}]$ such that $(A[x_{0},\ldots,x_{n}] / I)_{d}$ is a
flat $A$-module for every $d \geq \mu$ and for every prime $p \subset A$,the
Hilbert polynomial of the $k(p)$-graded algebra $(A[x_{0},\ldots,x_{n}]/I)
\otimes_{A} k(p)$ is equal to $\mu$.

By \cite{ueno}[Lem.7.51, p.55] and the bijection \eqref{bij2} introduced in
subsection \ref{grassrepr}, for $d \geq \mu$,
$\mathbf{Hilb}^{\mu}_{\mathbb{P}^{n}}(X)$ is the set of $A$-submodules
$I_{d}$ of $A[x_{0},\ldots,x_{n}]_{d}$ such that $A[x_{0},\ldots,x_{n}]_{d} /
I_{d}$ and $A[x_{0},\ldots,x_{n}]_{d+1} / I_{d+1}$ are locally free of rank
$\mu$, where  $I_{d+1}=A[x_{0},\ldots,x_{n}]_{1}.I_{d}$.

We can rewrite it as the set of $\epsilon$ locally free sheaves of modules of
rank $\mu$  on $X$ along with
$g:=\mathbb{K}[x_{0},\ldots,x_{n}]_{d}\otimes_{\mathbb{K}}\mathcal{O}_{X}
\rightarrow \epsilon \rightarrow 0$ such that $A[x_{0},\ldots,x_{n}]_{d+1} /
I_{d+1}$ is locally free of rank $\mu$, where  $I_{d+1} =
A[x_{0},\ldots,x_{n}]_{1}.\mathrm{Ker}(g)$.

Thus, we get a morphism of functors for $d \geq \mu$
$$
\Phi:=\mathbf{Hilb}^{\mu}_{\PP^{n}} \longrightarrow \mathbf{Gr}^{\mu}_{S_{d}^{*}}
$$
and a morphism of schemes 
$$
\phi:=\mathrm{Hilb}^{\mu}(\mathbb{P}^{n}) \longrightarrow \mathrm{Gr}^{\mu}(S_{d}^{*}).
$$

\begin{theorem}\label{fini}
 
The morphism $\phi$ is a closed immersion whose equations are \eqref{eq:hilbert:scheme} and \eqref{eq2:hilbert:scheme}. Equivalently, we have the following commutative diagram:
$$
\xymatrix{ \mathrm{Hilb}^{\mu}(\mathbb{P}^{n}) \ar[rr]^{\phi} \ar[rd]^{\sim} &&  \mathrm{Gr}^{\mu}(S_{d}^{*}) = \mathbf{Proj}(\mathbb{K}[\wedge^{\mu}S_{d}^{*}]/(\#)) \\ & \mathbf{Proj}(\mathbb{K}[\wedge^{\mu}S_{d}^{*}]/(\#,\eqref{eq:hilbert:scheme},\eqref{eq2:hilbert:scheme})) \ar[ru]^{\iota}}
$$
where $\iota$ is the natural closed immersion from
$\mathbf{Proj}(\mathbb{K}[\wedge^{\mu}S_{d}^{*}]/(\#,\eqref{eq:hilbert:scheme},\eqref{eq2:hilbert:scheme}))$
to $\mathbf{Proj}(\mathbb{K}[\wedge^{\mu}S_{d}^{*}]/(\#))$.
\end{theorem}
\begin{proof}
Let us prove that $\mathbf{Proj}(\mathbb{K}[\wedge^{\mu}S_{d}^{*}]/(\#,\eqref{eq:hilbert:scheme},\eqref{eq2:hilbert:scheme}))$ represents the Hilbert functor. Again we can reduce to the case of affine schemes $X=\mathbf{Spec}(A)$ with $A$ a noetherian $\mathbb{K}$ algebra. \\

Consider an element $(\epsilon,g)$ of $\mathbf{Hilb}^{\mu}_{\mathbb{P}^{n}}(X)$ with $\epsilon$ locally free sheaf of rank $\mu$ on $X$ and
$$
g: \mathbb{K}[x_{0},\ldots,x_{n}]_{d}\otimes_{\mathbb{K}}\mathcal{O}_{X} \rightarrow \epsilon \rightarrow 0.
$$
Let $I_{d}$ be the $A$ submodule of $A[x_{0},\ldots,x_{n}]_{d}$ such that
$\widetilde{I_{d}}=\text{Ker}(g)$ (thus $\epsilon$ is the quasi coherent
sheaf associated to $A[x_{0},\ldots,x_{n}]_{d}/I_{d}$). Let $I_{d+1}$ be the
submodule of $A[x_{0},\ldots,x_{n}]_{d+1}$ equal to
$A[x_{0},\ldots,x_{n}]_{1}.I_{d}$. By By proposition \ref{prop:1.5},
$A[x_{0},\ldots,x_{n}]_{d+1}/I_{d+1}$ is locally free of rank $\mu$. As we did
in the case of the Grassmannian functor, let $s_{\alpha} \in \epsilon(X)$
for $\alpha \in \mathbb{N}^{n}$ such that $|\alpha| = d$, be the image of the
monomial $x^{\alpha} \in S_{d}$ by $g$. For
$B=(x^{\alpha_{1}},\ldots,x^{\alpha_{\mu}})$ a family of $\mu$ monomials of
degree $d$ in $S_{d}$, we will denote by $p_{B} \in \wedge^{\mu} \epsilon(X)$
the global section  $p_{\alpha_{1},\ldots,\alpha{\mu}} = s_{\alpha_{1}}\wedge
\cdots \wedge s_{\alpha_{\mu}}$ (see lemma \ref{grass}).

We already know from theorem \ref{grassm}  that  $(p_{B})$, for  families $B$
of $\mu$ ordered monomials of degree $d$ in $A[x_{0},\ldots,x_{n}]_{d}$ (for
some monomial ordering $<$), satisfy the Pl\"ucker relations $(\#)$.

For all primes $p$ of $A$, let $(I_{p})_{d} = I_{d} \otimes A_{p}$ and
$(I_{p})_{d+1} = I_{d+1} \otimes A_{p}$. Then,
$A_{p}[x_{0},\ldots,x_{n}]_{d}/(I_{p})_{d}$ is the fiber of $\epsilon$ at $p \in
\mathbf{Spec}(A)$. Thus it a free $A_{p}$-module of rank $\mu$. By
definition, $A_{p}[x_{0},\ldots,x_{n}]_{d+1}/(I_{p})_{d+1}$ is also a free
$A_{p}$-module of rank $\mu$.

One can easily check that the image of the global sections $(p_{B})$, in the
stalk $\wedge^{\mu} \epsilon(X)_{p} = A_{p}$ of $\wedge^{\mu} \epsilon(X)$ at
$p \in \mathbf{Spec}(A)$, are exactly the coordinates $(\Delta_{B})$ of
$\Delta=T_{d}/(I_{p})_{d}$ introduced in subsection \ref{sec:2}. Thus, using
one side of the equivalence of theorem \ref{equi}, $(p_{B})$ satisfy
equations \eqref{eq:hilbert:scheme} and \eqref{eq2:hilbert:scheme} at any $p
\in \mathbf{Spec}(A)$. Thus, $(p_{B})$ satisfies globally equations
\eqref{eq:hilbert:scheme} and \eqref{eq2:hilbert:scheme} and we get a morphism
$$
X \longrightarrow \mathbf{Proj}(\mathbb{K}[\wedge^{\mu}S_{d}^{*}]/(\#,\eqref{eq:hilbert:scheme},\eqref{eq2:hilbert:scheme})).
$$
We deduce a morphism of contravariant functors
$$
\Psi:= \mathbf{Hilb}^{\mu}_{\PP^{n}} \longrightarrow \mathbf{Hom}(-,\mathbf{Proj}(\mathbb{K}[\wedge^{\mu}S_{d}^{*}]/(\#,\eqref{eq:hilbert:scheme},\eqref{eq2:hilbert:scheme}))).
$$\
\\
% \mathbf{Hilb}^{\mu}_{\mathbb{P}^{n}}(X)
$\Psi(X)$ is injective because, by definition, $\Phi(X)$ is injective and the following diagram is commutative:
$$
\xymatrix{ \mathbf{Hilb}^{\mu}_{\mathbb{P}^{n}}(X) \ar[rr]^{\Phi(X)} \ar[rd]^{\Psi(X)} &&  \mathbf{Gr}^{\mu}_{S_{d}^{*}}(X)  \\ & \mathbf{Hom}(X,\mathbf{Proj}(\mathbb{K}[\wedge^{\mu}S_{d}^{*}]/(\#,\eqref{eq:hilbert:scheme},\eqref{eq2:hilbert:scheme}))) \ar[ru]^{\iota}}.
$$
\
\\
Let us prove now that $\Psi(X)$ is surjective.\\
 Given an element of
 $\mathbf{Hom}(X,\mathbf{Proj}(\mathbb{K}[\wedge^{\mu}S_{d}^{*}]/(\#,\eqref{eq:hilbert:scheme},\eqref{eq2:hilbert:scheme})))$,
 we get (through its image in $\mathbf{Gr}^{\mu}_{S_{d}^{*}}(X)$) an element
 $(\epsilon,g)$ of $\mathbf{Hilb}^{\mu}_{\mathbb{P}^{n}}(X) $ with $\epsilon$
 locally free sheaf of rank $\mu$ on $X$ and 
$$
g: \mathbb{K}[x_{0},\ldots,x_{n}]_{d}\otimes_{\mathbb{K}}\mathcal{O}_{X} \rightarrow \epsilon \rightarrow 0.
$$
Using the same notations as above, for all primes $p \in \mathbf{Spec}(A)$, as $(p_{B})$ satisfy \eqref{eq:hilbert:scheme} and \eqref{eq2:hilbert:scheme}, $(\Delta_{B})$ also satisfy \eqref{eq:hilbert:scheme} and \eqref{eq2:hilbert:scheme}. Using the other side of the equivalence of theorem \ref{equi}, we get that $A_{p}[x_{0},\ldots,x_{n}]_{d+1}/(I_{p})_{d+1}$ is locally free of rank $\mu$ for all primes $p \in \mathbf{Spec}(A)$. Thus, $A[x_{0},\ldots,x_{n}]_{d+1}/I_{d+1}$ is locally free of rank $\mu$ and $(\epsilon,g)$ belongs to $\mathbf{Hilb}^{\mu}_{\mathbb{P}^{n}}(X) $.\\

Finally $\Psi(X)$ is a bijection and $\Psi$ is an isomorphism of contravariant functors. We conclude $\mathbf{Proj}(\mathbb{K}[\wedge^{\mu}S_{d}^{*}]/(\#,\eqref{eq:hilbert:scheme},\eqref{eq2:hilbert:scheme}))$ represents the Hilbert functor.
\end{proof}

\section{Tangent space}\label{sec:3}

Our objective in this section is to determine the tangent space at a
$\mathbb{K}$-rational point $I_{0}$ of the Hilbert scheme $\HilbPPn$. From
Theorem \ref{fini}, $\HilbPPn$ is a projective scheme whose equations are the
equations \eqref{eq:hilbert:scheme}, \eqref{eq2:hilbert:scheme} and the
Pl\"ucker relations $(\#)$.

For $\mathbf{u}$ in $S_{1}$, let $H_{\mathbf{u}}$ be the open subscheme of
$\HilbPPn$ associated to the open subfunctor $\Hb_{\mathbf{u}}$ introduced in
definition \ref{hudef}.  Let $\mathbf{v} \in S_{1}$ be a linear form such
that the open subscheme $H_{\textbf{v}}$ contains $I_{0}$ (ie. $\mathbf{v}$
is not a zero divisor of $I_{0}$).  After a change of coordinates
($\textbf{v} = x_{0}$) we can assume this open subset is of the form
$H_{\textbf{x}_{0}}$. Let $B \in \mathcal{B}_{d}$ (see lemma \ref{lemme}) be
a family of $\mu$ monomials in $S_{d}$ (for $d \geq \mu$) connected to 1.
Let $H^{B}_{x_{0}}$ be the open affine subscheme associated to the open
subfunctor $\Hb^{B}_{x_{0}}$ of the Hilbert scheme $\HilbPPn$ (see definition
\ref{hx0b}). Using proposition \ref{H} and lemma \ref{lemme},
$H^{B}_{\textbf{x}_{0}}$ is the affine scheme associated to the affine
variety
$$
\mathfrak{H}_{\underline{B}} \assign \{ \tmmathbf{z} \in
  \kk^{\mu \times N} ; M_{x_i}(\tmmathbf{z}) \circ M_{x_j}(\tmmathbf{z}) - M_{x_j}(\tmmathbf{z}) \circ M_{x_i}(\tmmathbf{z})
  = 0, 1 \leqslant i < j \leqslant n\}.$$ 
The system of coordinates of this variety is the set of parameters $\tmmathbf{z}=(z_{\alpha,\beta})_{\alpha\in \partial B,
  \beta\in B}$ such that 
$$ 
h_{\alpha}^{0}(\xbb) = \xbb^{\alpha} -\sum _{\beta\in B} z_{\alpha,\beta}\,
\xbb^{\beta} 
$$
is a border basis of $\underline{I_{0}}$ for $B$. 
%where $(z_{\alpha,\beta})_{\alpha\in \partial B, \beta\in B}$ = $({\Delta_{x_{0} B^{[x_{0}b_{\beta}|b_{\alpha}]}}\over \Delta_{x_{0}B}})_{\alpha\in \partial B, \beta\in B}$ (the rest of the coordinates $\Delta_{B'}$ for $B'$ not of the form $x_{0} B^{[x_{0}b_{\beta}|b_{\alpha}]}$, can by expressed as polynomial functions of $({\Delta_{x_{0} B^{[x_{0}b_{\beta}|b_{\alpha}]}}\over \Delta_{x_{0}B}})_{\alpha\in \partial B, \beta\in B}$). 
Then, the equations of $\HilbPPn$ in this system of coordinates reduce to the commutation relations:
$$
 M_{i}^{0}(\tmmathbf{z})\,M_{j}^{0}(\tmmathbf{z})-M_{j}^{0}(\tmmathbf{z})\,M_{i}^{0}(\tmmathbf{z})=0,
$$
where $M_{i}^{0}(\tmmathbf{z})$ is the operator of multiplication by $x_{i}$ in the
basis $B$ modulo the affine ideal $\underline{I_{0}}$.
We will compute the tangent space of the Hilbert scheme using the
previous system of coordinates.

By definition, the tangent space of $\HilbPPn$ at $I_{0}$ is the set of vectors
$ \hb^1= (h_{\alpha, \beta}^1)$ such that line 
 $h_{\alpha}^{\varepsilon}(\xb) = h_{\alpha}^0(\xb) + \varepsilon h_{\alpha}^1(\xb)$
intersects $\HilbPPn$  with multiplicity $\ge 2$,  where $h_{\alpha}^1 ( \xbb) \assign
\sum_{\beta \in B} h_{\alpha, \beta}^1 \xbb^{^{\beta}}$.

Substituting in the commutation relations, we obtain
\begin{eqnarray*}
\lefteqn{M_{x_i}^{\varepsilon} \circ M_{x_j}^{\varepsilon} - M_{x_j}^{\varepsilon}
  \circ M_{x_i}^{\varepsilon}}\\ & = & (M_{x_i}^0 \circ M_{x_j}^0 - M_{x_j}^0
  \circ M_{x_i}^0)\\
  &  & + \varepsilon (M_{x_i}^1 \circ M_{x_j}^0 + M_{x_i}^0 \circ M_{x_j}^1 -
  M_{x_j}^1 \circ M_{x_i}^0 - M_{x_j}^0 \circ M_{x_i}^1) + \mathcal{O}
  (\varepsilon^2)\\
  & = & \varepsilon (M_{x_i}^1 \circ M_{x_j}^0 + M_{x_i}^0 \circ M_{x_j}^1 -
  M_{x_j}^1 \circ M_{x_i}^0 - M_{x_j}^0 \circ M_{x_i}^1) + \mathcal{O}
  (\varepsilon^2) .
\end{eqnarray*}
where  $M_{x_i}^{\varepsilon} = M_{x_i}^0 +
\varepsilon M_{x_i}^1$, $M_{x_i}^0$ is the
operator of multiplication by $x_i$ in $\Ac^0$ and $M_{x_i}^1$ is
linear in $\tmmathbf{h}^1$.
We deduce the linear equations in $\tmmathbf{h}^1$ defining the tangent
space of $\HilbPPn$ at $I_{0}$:
\begin{equation}
  M_{x_i}^1 \circ M_{x_j}^0 + M_{x_i}^0 \circ M_{x_j}^1 - M_{x_j}^1 \circ
  M_{x_i}^0 - M_{x_j}^0 \circ M_{x_i}^1 = 0\ (1 \leqslant i < j \leqslant n) .
  \label{eq:commute:der}
\end{equation}
\begin{definition}
  Let $\tmmathbf{h}^0 \assign (h_{\alpha}^0)_{\alpha \in \partial B}$ be a
  border basis for $B$. We denote by $T_{\tmmathbf{h}_0}$, the set of
  $\tmmathbf{h}^1 = (h_{\alpha}^1)_{\alpha \in \partial B}$ with $h_{\alpha}^1
  \in \langle B \rangle$, which satisfies the linear equations
  (\ref{eq:commute:der}).
\end{definition}

In the following, we will also denote by $H^0 : \langle B^+ \rangle
\rightarrow \langle B^+ \rangle$ the linear map such that for $\beta \in B$,
$H^0 ( \xbb^{\beta}) = 0$ and for $\alpha \in \partial B$,
$H^0 ( \xbb^{\alpha}) = h_{\alpha}^0$. 
We also denote by $N_{0}: \langle B^+ \rangle
\rightarrow \langle B \rangle$  the normal form modulo $\Ic^{0}$, that is
the projection of $\langle B^+ \rangle$ on $\langle B \rangle$ along $\langle
h^{0}_{\alpha}\rangle$.
By construction, for any $p =
\sum_{\alpha \in B^+} \lambda_{\alpha} \xbb^{\alpha} \in \langle B^+
\rangle$, $H^0 (p) = \sum_{\alpha \in \partial B} \lambda_{\alpha}
h_{\alpha}^0$ and we have $N^0 + H^0 = \tmop{Id}_{\langle B^+ \rangle}$.
Similarly, we also denote by $H^1 : \langle B^+ \rangle \rightarrow \langle B
\rangle$ the map defined by $H^1 ( \xbb^{\beta}) = 0$ if
$\xbb^{\beta} \in B$, $H^1 ( \xbb^{\alpha}) = h^1_{\alpha}$ if
$\alpha \in \partial B$. By construction, for all $m \in B$, \ $M_i^1 (m) =
H^1 (x_i m)$.

\begin{theorem}
  Let $I_0 \in H_{x_{0}}^{B}$ be an ideal, with the border relations 
  $\tmmathbf{h}^0 \assign (h_{\alpha}^0)_{\alpha \in \partial B}$ for the
  basis $B$ of $\Ac^0 = R / \Ic_0$, where $\Ic_{0}=\underline{I_{0}}$. Then
  \begin{eqnarray*}
    \phi : T_{\tmmathbf{h}^0} & \rightarrow & \tmop{Hom}_R ( \Ic_0, R
    / \Ic_0)\\
    \tmmathbf{h}_{}^1 & \rightarrow & \phi ( \tmmathbf{h}^1) : h_{\alpha}^0
    \mapsto h_{\alpha}^1
  \end{eqnarray*}
  is an isomorphism of $\kk$-vector spaces.
\end{theorem}

\begin{proof}
  We first prove that $\phi ( \tmmathbf{h}^1)$ is well-defined, ie. if $g =
  \sum_i u_{\alpha} h_{\alpha}^0 = \sum_{\alpha} u_{\alpha}' h_{\alpha}^0 \in
  \Ic_0$ with $u_{\alpha}, u_{\alpha}' \in R$, then $\sum_{\alpha}
  u_{\alpha} h_{\alpha}^1 = \sum_{\alpha} u_{\alpha}' h_{\alpha}^1$ in $R /
  \Ic_0 .$ In other words, if $\sum_{\alpha} v_{\alpha} h_{\alpha}^0
  = 0$ in $R$, then $\sum_{\alpha} v_{\alpha} h_{\alpha}^1 \equiv 0$ in $R /
  \Ic_0$. 
  By {\cite{MT:08:tcs}[Theorem 4.3]}, the syzygies of the border basis
  elements $\tmmathbf{h}^0 \assign (h_{\alpha}^0)$ are generated by the
  commutation polynomials:
  \[ x_i H^0 (x_{i'} m) - x_{i'} H^0 (x_i m) + H^0 (x_i N^0 (x_{i'} m)) - H^0
     (x_{i'} N^0 (x_i m)), \]
  for all $m \in B$, $1 \leqslant i < i' \leqslant n$. Let us prove that these
  relations are also satisfied modulo $\Ic_0$, if we replace $H^0$ by
  $H^1$. As $\tmmathbf{h}^1 = (h_{\alpha}^1) \in T_{\tmmathbf{h}_0}$, we have
  \begin{eqnarray*}
    0 & = & M_{x_i}^0 \circ M_{x_{i'}}^1 (m) - M_{x_{i'}}^0 \circ M_{x_i}^1
    (m) + M_{x_i}^1 \circ M_{x_{i'}}^0 (m) - M_{x_{i'}}^1 \circ M_{x_i}^0
    (m)\\
    & = & N^0 (x_i H^1 (x_{i'} m)) - N^0 (x_{i'} H^1 (x_i m)) + H^1 (x_i N^0
    (x_{i'} m)) - H^1 (x_{i'} N^0 (x_i m))\\
    & = & x_i H^1 (x_{i'} m) - H^0 (x_i H^1 (x_{i'} m))\\
    &  & - x_{i'} H^1 (x_i m) + H^0 (x_{i'} H^1 (x_i m))\\
    &  & + H^1 (x_i N^0 (x_{i'} m)) - H^1 (x_{i'} N^0 (x_i m))\\
    & \equiv & x_i H^1 (x_{i'} m) - x_{i'} H^1 (x_i m) + H^1 (x_i N^0 (x_{i'}
    m)) - H^1 (x_{i'} N^0 (x_i m)) \tmop{modulo} \Ic_0 .
  \end{eqnarray*}
  This proves that the generating syzygies are mapped by $\phi ( \tmmathbf{h}^1)$
  to $0$ in $R / \Ic_0$ and thus the image by $\phi (
  \tmmathbf{h}^1)$ of any syzygy of $\tmmathbf{h}^0$ is $0$, that is,  $\phi (
  \tmmathbf{h}^1)$ is a well-defined element of $\tmop{Hom}_R (\Ic_0, R/\Ic_0)$.
  
  Conversely, let us prove that if $\psi_0 \in \tmop{Hom}_R ( \Ic_0,
  R / \Ic_0)$, then $\tmmathbf{h}^1 : = (\psi_0 (h^0_{\alpha})) \in
  T_{\tmmathbf{h}_0}$. As $\psi_0 \in \tmop{Hom}_R ( \Ic_0, R /
  \Ic_0)$, the syzygies of $\tmmathbf{h}^0$ are mapped by $\psi_0$ to
  $0$. Thus, for all $m \in B$, $1 \leqslant i < i' \leqslant n$,
  
  $\begin{array}{lll}
    0 & \equiv & \psi_0 (x_i H^0 (x_{i'} m) - x_{i'} H^0 (x_i m) + H^0 (x_i
    N^0 (x_{i'} m)) - H^0 (x_{i'} N^0 (x_i m)))\\
    & \equiv & x_i H^1 (x_{i'} m) - x_{i'} H^1 (x_i m) + H^1 (x_i N^0 (x_{i'}
    m)) - H^1 (x_{i'} N^0 (x_i m)) \tmop{modulo} \Ic_0 .
  \end{array}$
  
  As $H^1 (p) \in \langle B \rangle$ and $N^0 (H^1 (p)) = H^1 (p)$ for all
  $p \in \langle B^+ \rangle$, we have
  
  $\begin{array}{lll}
    0 & = & N^0 (x_i H^1 (x_{i'} m)) - N^0 (x_{i'} H^1 (x_i m)) + H^1 (x_i N^0
    (x_{i'} m)) - H^1 (x_{i'} N^0 (x_i m))\\
    & = & M^0_i \circ M^1_{i'} (m) - M^0_{i'} \circ M^1_i (m) + M^1_i \circ
    M^0_{i'} (m) - M^1_{i'} \circ M^0_i (m),
  \end{array}$
  which proves that $\tmmathbf{h}^1 \in T_{\tmmathbf{h}^0}$.
\end{proof}

We can notice that the tangent space of the variety $\HilbPPn$ locally defined by the
equations \eqref{eq:commute:der} is also isomorphic to $\tmop{Hom}_R ( \Ic_{0}/\Ic^{2}_0, R
    / \Ic_0)$. Our construction gives a new (simple) proof of this well known
    result  \cite{sernesi}[p. 217].

\bigskip
\appendix
The results in the following appendix parts can be considered as
``classical'', though not necessarily explicit in the literature. They are
recalled here for the sake of completeness.
\section{Representable functors}\label{representable}

We consider the category $\mathcal{C}$ of noetherian schemes over
$\mathbb{K}$. Let $\mathcal{C}^{a}$ be the category of affine noetherian schemes over $\mathbb{K}$. Let $\mathbb{P}^{n}$ be the projective scheme
$\mathbf{Proj}(S)$. For the notions of presheaf, sheaf and scheme,  see
\cite{hartshorne}[Chap. II]. The objective of this section is to find
conditions to the representation of contravariant functors from the category
of schemes to the category of sets. Most of the material used in this section
comes from appendix E of \cite{sernesi}.

\begin{proposition}\label{represente}
 Let $F$ be a contravariant functor from the category $\mathcal{C}$ to the category of Sets. Suppose that:\\
- $F$ is a sheaf\\
- $F$ admits an open covering of representable functors,\\
then $F$ is also representable.
\end{proposition}
\begin{proof}
See appendix E in \cite{sernesi}[Prop.E.10, p.318] 
\end{proof}

\begin{proposition}\label{functor}
 Let $F$ be a contravariant functor from $\mathcal{C}$ to the category of
 Sets and $G$ a subfunctor of $F$. Assume that for every affine scheme $X$ in
 $\mathcal{C}$ and every morphism of functors: 
$$
\mathbf{Hom}(-,X) \rightarrow F
$$
the functor $H := \mathbf{Hom}(-,X) \times_{F} G$ restricted to the category of affine noetherian schemes over $\mathbb{K}$ is represented by an open subscheme of $X$. Then $G$ is an open subfunctor of $F$ (in $\mathcal{C}$).
\end{proposition}
\begin{proof}
 Let $X$ and $Y$ be objects of $\mathcal{C}$. Let $(U_{i})_{i \in I}$ be any affine covering of $X$. Consider the morphism of functors from $\mathbf{Hom}(-,X)$ to $F$ given by an element $\lambda \in F(X)$ (see appendix E in \cite{sernesi}[Lem.E.1, p.313]). Then, the contravariant functor $H := \mathbf{Hom}(-,X) \times_{F} G$ is given by:
$$
H(Y) = \{ \phi \in \mathbf{Hom}(Y,X) | F(\phi)(\lambda) \in G(Y) \subset F(Y) \}
$$
Let $(V_{i,j})_{j \in J}$ be an affine covering of $\phi^{-1}(U_{i}) \subset Y$ for all $i$. Let $\phi_{i,j}$ be the restriction of $\phi$ to $V_{i,j}$:
$$
\phi_{i,j}: V_{i,j} \rightarrow U_{i}.
$$
As $F$ and $G$ are sheaves, $F(\phi)(\lambda) \in G(Y)$ if and only if $F(\phi_{i,j})(\lambda_{i}) \in G(V_{i,j})$ (where $\lambda_{i} \in F(U_{i})$ is the restriction of $\lambda \in F(X)$ to $F(U_{i})$).\\
As $G$ is an open subfunctor of $F$ in the category of affine schemes, there exists an open subscheme $\Omega_{i}$ of the  affine scheme $U_{i}$ such that:\\

\begin{equation}\label{equivalence}
F(\phi)(\lambda) \in G(Y) \Leftrightarrow  F(\phi_{i,j})(\lambda_{i}) \in G(V_{i,j}) \Leftrightarrow \phi_{i,j} \text{ factors through }\Omega_{i}:
\end{equation}

$$
\xymatrix{ V_{i,j} \ar[rr]^{\phi_{i,j}} \ar[rd] && U_{i}   \\ & \Omega_{i} \ar[ru] }
$$\\

If $\phi \in \mathbf{Hom}(Y,X)$ belongs to $H(Y)$, $F(\phi)(\lambda) \in G(Y)$ by definition. Thus, from the previous equivalence \eqref{equivalence}, $\phi$ factors through $\Omega := \bigcup_{i} \Omega_{i} \subset X$:
$$
\xymatrix{ Y \ar[rr]^{\phi} \ar[rd] && X   \\ & \Omega=\bigcup\Omega_{i} \ar[ru] }
$$
Reciprocally, if $\phi$ factors through $\Omega$, consider $(W_{i,j})_{j \in J}$ an open affine covering of $\phi^{-1}(\Omega_{i}) \subset Y$ for all $i \in I$. We denote now by $\phi_{i,j}$ the restriction of $\phi$ to $W_{i,j}$:
$$
\xymatrix{ W_{i,j} \ar[rr]^{\phi_{i,j}} \ar[rd] && U_{i}  \\ & \Omega_{i} \ar[ru] }
$$
This commutative diagram and \eqref{equivalence} imply that
$F(\phi_{i,j})(\lambda_{i}) \in F(W_{i,j})$ belongs to $G(W_{i,j})$ for all
$i,j$. It implies (as $F$ and $G$ are sheaves) that $F(\phi)(\lambda) \in
F(Y)$ belongs to $G(Y)$ and thus $\phi$ belongs to $H(Y)$.

Thus, the functor $H$ is isomorphic $\mathbf{Hom}(-,\Omega)$, i.e $H$ is
represented by the open subscheme $\Omega$ of $X$. We conclude that $G$ is an open subfunctor of $F$ (in $\mathcal{C}$).
\end{proof}
\begin{proposition}\label{affine}
 Let $F$ be a contravariant functor from $\mathcal{C}$ to the category of
 Sets. $F$ is represented by the scheme $X$ if and only if the functors $F$
 and $\mathbf{Hom}(-,X)$ are isomorphic in the category of affine schemes
 over $\mathbb{K}$. 
\end{proposition}
\begin{proof}
 This is a straightforward consequence of the fact that every scheme has a
 topological basis which consists of open affine subschemes. 
\end{proof}

\section{The Grassmannian}\label{grassmannian}
The objective of this section is to present a construction of the
Grassmannian as a scheme representing a contravariant functor. Most of the
material used for this construction comes from \cite{sernesi}[Chap.4.3.3,
  p.209].

\begin{theorem}\label{grassm}
For all $\mathbb{K}$-vector spaces $V$ of finite dimension $N$  and integers $n\leq N$, the Grassmannian functor $n$ of $V$ is representable. It is represented by a projective scheme we will denote $\mathrm{Gr}^{n}(V)$:
$$
 \mathrm{Gr}^{n}(V) \sim \mathbf{Proj}(\mathbb{K}[\wedge^{n}V]/(\#)).
$$ 
where $(\#)$ is the ideal generated by the Pl\"ucker relations. 
\end{theorem}
\begin{proof}
By definition, the $n$ Grassmannian functor of $V$ is:
$$
X \longrightarrow \{ V^{*} \otimes_{\mathbb{K}} \mathcal{O}_{X} \rightarrow \epsilon \rightarrow 0 \ | \ \epsilon \text{ is a locally free sheaf of rank } n \text{ of } \mathcal{O}_{X} \}.
$$
Let 
$$
g := V^{*} \otimes_{\mathbb{K}} \mathcal{O}_{X} \rightarrow \epsilon \rightarrow 0
$$
be an element of $\mathbf{Gr}^{n}_{V}(X)$. Let $\wedge$ denote the exterior product. Then we have
$$
\wedge^{n} g := \wedge^{n} V^{*} \otimes_{\mathbb{K}} \mathcal{O}_{X} \rightarrow \wedge^{n} \epsilon = \mathcal{L} \rightarrow 0
$$
where $\mathcal{L}$ is an invertible sheaf. Let $(e_{0},\ldots,e_{N})$ be a basis of $V$. Let $s_{i} \in \epsilon (X)$ be the image of $e_{i}$ by $g$ and $p_{i_{1},\ldots,i_{n}}:= s_{i_{1}}\wedge\ldots\wedge s_{i_{n}} \in \mathcal{L}(X)$. $(p_{i_{1},\ldots,i_{n}})$ satisfies the well-known Pl\"ucker relations 
$$
(\#) \ \ \sum_{\lambda=1,\ldots,n+1} p_{i_{1},\ldots,i_{n-1},j_{\lambda}} \otimes p_{i_{1},\ldots,\hat{j_{\lambda}},\ldots,i_{n}} = 0.
$$
Thus, by \cite{EGA2}[Prop.4.2.3, p.73] we have a morphism:
$$
X \longrightarrow \mathbf{Proj}(\mathbb{K}[\wedge^{n}V]/(\#))
$$ 
Thus we constructed a morphism of functors from the Grassmannian functor $\mathbf{Gr}^{n}_{V}$ to the functor $\mathbf{Hom}(-,\mathbf{Proj}(\mathbb{K}[\wedge^{n}V]/(\#)))$:
$$
\Phi := \mathbf{Gr}^{n}_{V} \longrightarrow \mathbf{Hom}(-,\mathbf{Proj}(\mathbb{K}[\wedge^{n}V]/(\#)))
$$
\begin{lemma}\label{grass}
 The morphism $\Phi$ is an isomorphism of functors. Thus the Grassmannian functor is represented by $\mathbf{Proj}(\mathbb{K}[\wedge^{n}V]/(\#))$.
\end{lemma}
\begin{proof}
From proposition \ref{affine}, we can reduce to affine schemes $X=\mathbf{Spec}(A)$. Let us prove that
$$
\Phi(X) := \mathbf{Gr}^{n}_{V}(X) \rightarrow \mathbf{Hom}(X,\mathbf{Proj}(\mathbb{K}[\wedge^{n}V]/(\#)))
$$
is a bijection.\\
Let locally free sheaves of rank $\mu$  $\epsilon$ and $\epsilon'$ together with surjective morphisms
$$
g=V^{*}\otimes \mathcal{O}_{X} \rightarrow \epsilon \rightarrow 0
$$
and 
$$
g'=V^{*}\otimes \mathcal{O}_{X} \rightarrow \epsilon ' \rightarrow 0
$$
be two elements of $\mathbf{Gr}^{n}_{V}(X)$. Let $(s_{i})$ and $(p_{i_{1},\ldots,i_{n}})$  (resp. $(s'_{i})$ and $(p'_{i_{1},\ldots,i_{n}})$) be the global sections of $\epsilon$ and $\mathcal{L}=\wedge^{n}\epsilon$ (resp. $\epsilon '$ and $\mathcal{L'}=\wedge^{n}\epsilon '$) introduced before. Assume $\Phi(X)(\epsilon,g) \in \mathbf{Hom}(X,\mathbf{Proj}(\mathbb{K}[\wedge^{n}V]/(\#)))$ and $\Phi(X)(\epsilon ',g') \in \mathbf{Hom}(X,\mathbf{Proj}(\mathbb{K}[\wedge^{n}V]/(\#)))$ are equal. Then there exists an isomorphism $\phi$ between $\mathcal{L}$ and $\mathcal{L'}$ such that the following diagram is commutative:
$$
\xymatrix{ \wedge^{n}V^{*}\otimes \mathcal{O}_{X} \ar[rr] \ar[rd] && \mathcal{L} \ar[ld]^{\phi} \\ & \mathcal{L'} }
$$
and such that $\phi(p_{i_{1},\ldots,i_{n}}) = p'_{i_{1},\ldots,i_{n}}$.\\
Consider the open subset $X_{p_{i_{1},\ldots,i_{n}}}$ where $p_{i_{1},\ldots,i_{n}} \neq 0$ (which is equal to $X_{p'_{i_{1},\ldots,i_{n}}}$ because $\phi$ is an isomorphism that sends $p_{i_{1},\ldots,i_{n}}$ on $p'_{i_{1},\ldots,i_{n}}$). On $X_{p_{i_{1},\ldots,i_{n}}}$ (resp. $X_{p'_{i_{1},\ldots,i_{n}}}$), $(s_{i_{1}},\ldots,s_{i_{n}})$ (resp. $(s'_{i_{1}},\ldots,s'_{i_{n}})$) is a basis of $\epsilon$ (resp. $\epsilon$ ') and we have 
$$
s_{j} = \sum_{k=1,\ldots,n} a_{j,k}.s_{i{_k}} \ (resp. \ s'_{j} = \sum_{k=1,\ldots,n} a'_{j,k}.s'_{i{_k}})
$$
with 
\begin{equation}\label{coordo}
a_{jk} = (-1)^{n-k}\frac{p_{i_{1},\ldots,\hat{i_{k}},\ldots,i_{n},j}}{p_{i_{1},\ldots,i_{n}}} = a'_{jk}.
\end{equation}
Thus, on $X_{p_{i_{1},\ldots,i_{n}}}$, we have the natural isomorphism $f_{i_{1},\ldots,i_{n}}$ from $\epsilon$ to $\epsilon '$ that sends $s_{i_{k}}$ to $s'_{i_{k}}$ for all $k \leq n$. Then, by equations \eqref{coordo}, the morphisms $(f_{i_{1},\ldots,i_{n}})$ patch together to form an isomorphism $f$ from $\epsilon$ to $\epsilon '$ such that:

$$
\xymatrix{ V^{*}\otimes \mathcal{O}_{X} \ar[rr]^{g} \ar[rd]^{g'} && \epsilon \ar[ld]^{f} \\ & \epsilon ' }
$$
is commutative. Thus 
$$
g=V^{*}\otimes \mathcal{O}_{X} \rightarrow \epsilon \rightarrow 0
$$
and 
$$
g'=V^{*}\otimes \mathcal{O}_{X} \rightarrow \epsilon ' \rightarrow 0
$$
are equal as elements of $\mathbf{Hom}(X,\mathbf{Proj}(\mathbb{K}[\wedge^{n}V]/(\#)))$. Thus $\Phi(X)$ is injective.\\

To prove $\Phi(X)$ is surjective, let 
$$
\phi:=\wedge^{n}V^{*}\otimes \mathcal{O}_{X} \rightarrow \mathcal{L} \rightarrow 0
$$
be an element of $\mathbf{Hom}(X,\mathbf{Proj}(\mathbb{K}[\wedge^{n}V]/(\#)))$ where $\mathcal{L}$ is an invertible sheaf on $X$ and such that $p_{i_{1},\ldots,i_{n}} := \phi(e_{i_{1},\ldots,i_{n}})$ satisfy the relations $(\#)$. On $X_{p_{i_{1},\ldots,i_{n}}}$ let $\epsilon_{i_{1},\ldots,i_{n}}$ be a free sheaf of rank $n$ with basis $(e_{i_{1}},\ldots,e_{i_{n}})$. Using relations  \eqref{coordo} and $(\#)$, we can glue the sheaves $(\epsilon_{i_{1},\ldots,i_{n}})$ to form a locally free sheaf $\epsilon$ of rank $n$ together with a surjective morphism
$$
g:=V^{*}\otimes \mathcal{O}_{X} \rightarrow \epsilon \rightarrow 0
$$
which satisfies 
$$
\Phi(X)(\epsilon,g) = (\mathcal{L,\phi}).
$$
Thus $\Phi(X)$ is surjective and $\Phi$ is an isomorphism of functors.
\end{proof}
From lemma \ref{grass}, the Grassmannian functor is represented by
$$
 \mathrm{Gr}^{n}(V) \sim \mathbf{Proj}(\mathbb{K}[\wedge^{n}V]/(\#)).
$$
\end{proof}

\section{Generic linear forms}\label{Generic}

The objective of this section is the to extend the notion of generecity in the case of polynomial rings over a field $\mathbb{K}$ to the case of polynomial rings over a $\mathbb{K}$-algebra.\\
In the following, $\mathbb{K}$ will denote a field of characteristic zero.

\begin{proposition}\label{generic}
 Let $A$ be a $\mathbb{K}$-algebra and $n$ be an integer. Let $P$ be a polynomial in $A[x_{1},\ldots,x_{n}]$ such that $P$ vanishes on generic values of $\mathbb{K}^{n}$, then $P= 0$.
\end{proposition}
\begin{proof}
 Let $\mathbf{P_{d}}$ be the following proposition: for all $m \in \NN$ and
 all polynomial $P \in A[x_{1},\ldots,x_{m}]$ of degree less or equal to $d$, if
 $P$ vanishes on generic bvalues of $\mathbb{K}^{m}$ then $P= 0$. We will prove by induction that $\mathbf{P_{d}}$ is true for all $d \geq 0$.\\
For $d = 0$, $\mathbf{P_{0}}$ is obviously true.\\
Assume $\mathbf{P_{k}}$ is true for all $k \leq d$, let us prove that $\mathbf{P_{d+1}}$ is true. Let $m$ be an integer and $P$ be a polynomial in $A[x_{1},\ldots,x_{m}]$ of degree less or equal to $d+1$ such that $P$ vanishes on generic values of $\mathbb{K}^{m}$. Let $U \subset \mathbb{K}^{m}$ be the zero set of $P$. Let $Q_{i} \in A[x_{1},\ldots,x_{m},y_{1},\ldots,y_{m}]$ be the polynomial given by
$$
Q_{i}(\mathbf{x},\mathbf{y}) = {P(\mathbf{x}) - P(\mathbf{y}) \over x_{i}- y_{i}}.
$$
As $P=(x_{i}-y_{i}).Q_{i}$, $Q_{i}$ vanishes on $V:= U \times U \setminus \{
(\mathbf{x},\mathbf{y}) | \ x_{i}-y_{i} = 0\} \subset \mathbb{K}^{2m}$. Thus
$Q_{i}$ vanishes on generic values of $\mathbb{K}^{2m}$ and is of degree less
or equal to $d$. By $\mathbf{P_{d}}$, $Q_{i}$ is equal to $0$ for all $1 \leq
i \leq m$. Thus all the partial derivatives
$\partial_{i}P=Q_{i}(\mathbf{x},\mathbf{x})$ of $P$ are equal to 0. As
$\mathbb{K}$ is of charasteristic zero, we conclude $P$ is equal to 
zero.\\ 
Thus $\mathbf{P_{d}}$ is true for all $d$ and $m$ in $\NN$. This proves the proposition \ref{generic}.
\end{proof}

\section{Zero dimensional algebra}
In this section, we recall why an ideal $I$ remains in
$\mathbf{Hilb}^{\mu}_{\mathbb{P}^{n}}$ by coefficient field extension and give
a characterization of non-zero divisibility for linear forms.

Let $k$ be a field of characteristic zero and $\overline{k}$ its algebraic
closure. Let $S=k[x_{0},\ldots,x_{n}]$ (resp. $\overline{S}:=S
\otimes_{k}\overline{k}$) be the polynomial ring in $n+1$ variables over $k$
(resp. $\overline{k}$). 
Recall that $\mathbf{Hilb}^{\mu}_{\mathbb{P}^{n}}(\mathbf{Spec}(k))$ (or simply $\mathbf{Hilb}^{\mu}_{\mathbb{P}^{n}}(k)$) is equal to the set of homogeneous saturated ideal of $S$ such $S/I$ has Hilbert polynomial equal to the constant $\mu$.\\
Given a point $P$ in the projective space $\mathbb{P}^{n}_{k}$ we denote by $m_{k,P}$ the homogeneous ideal of $k[x_{0},\ldots,x_{n}]$ generated by
$$
\{  Q \text{ homogeneous polynomial in } k[x_{0},\ldots,x_{n}] | \ Q(P_{i}) = 0\}.
$$
Finally, denote by $m_{k}$ the homogeneous ideal of $k[x_{0},\ldots,x_{n}]$ generated by 
$$
\{P \text{ homogeneous polynomial in } k[x_{0},\ldots,x_{n}] \text{ of degree}\geq 1  \}.
$$
\begin{definition}
 Let $I$ be a homogeneous ideal of $\mathbf{Hilb}^{\mu}_{\mathbb{P}^{n}}(k)$. Then, $\overline{I}$ is the ideal of $\overline{S}$ given by
$$
\overline{I} := I \otimes_{k} \overline{k}.
$$
\end{definition}

\begin{proposition}\label{ibarre}
 Let $I \subset S$ be a homogeneous ideal of $\mathbf{Hilb}^{\mu}_{\mathbb{P}^{n}}(k)$. Then, one has 
$$
\overline{I}\cap S  = I
$$
\end{proposition}
\begin{proof}
 First, one has that $I \subset \overline{I}\cap S$. Then, tensoring by $\overline{k}$, one has that $\overline{I} \subset (\overline{I}\cap S)\otimes\overline{k} \subset \overline{I}$. Thus $\overline{I} =(\overline{I}\cap S)\otimes\overline{k}$. Then, looking at the dimensions for all degree $d \leq 1$, one has that
$$
\mathrm{dim}_{k}(S_{d}/(\overline{I}\cap S)_{d}) = \mathrm{dim}_{\overline{k}}(S_{d}/(\overline{I}\cap S)_{d}\otimes\overline{k}) =\mathrm{dim}_{\overline{k}}(S_{d}/I_{d}\otimes\overline{k}) = \mathrm{dim}_{k}(S_{d}/I_{d}).
$$
 We conclude that
$$
I = \overline{I}\cap S.
$$
\end{proof}

\begin{corollary}\label{ibarrecor}
 Let $I \subset S$ be a homogeneous ideal of $\mathbf{Hilb}^{\mu}_{\mathbb{P}^{n}}(k)$. Then $\overline{I}$ belongs to $\mathbf{Hilb}^{\mu}_{\mathbb{P}^{n}}(\overline{k})$.
\end{corollary}
\begin{proof}
We just need to prove that $\overline{I}$ is saturated (i.e
$\overline{I}:m_{\overline{k}}= \overline{I}$). One has that $\overline{I}$
is saturated if and only if $m_{\overline{k}}$ is not a prime associated to
$\overline{I}$. Assume $\overline{I}$ is not saturated. From the Nullstellensatz theorem, $\overline{I}$ has a reduce primary decomposition of the form
$$
\overline{I} = \bigcap_{i} q_{i} \cap q
$$ 
with $q_{i}$ homogeneous $m_{\overline{k},P_{i}}$-primary ideal with $P_{i}$ a point in the projective space $\mathbb{P}^{n}_{\overline{k}}$, and $q$ a homogeneous $m_{\overline{k}}$-primary ideal. From proposition \ref{ibarre}, we have 
$$
I = \overline{I}\cap S.
$$ 
Thus, one has
$$
I = \bigcap_{i} q_{i}\cap S \bigcap q\cap S
$$
with $q_{i}\cap S$ homogeneous $m_{\overline{k},P_{i}}\cap S$-primary ideal and $q\cap S$ homogeneous $m_{\overline{k}}\cap S = m_{k}$-primary ideal. As $I$ is saturated this is impossible. Thus $\overline{I}$ is saturated.
\end{proof}

\begin{proposition}\label{vanish}
 Let $I$ be a homogeneous ideal in $\mathbf{Hilb}^{\mu}_{\mathbb{P}^{n}}(k)$
 and $u$ a linear form in $S_{1}$. Then $(I:u)=I$ if and only if $u$ does not
 vanish at any point defined by $\overline{I}$ in
 $\mathbb{P}^{n}_{\overline{k}}$. 
\end{proposition}
\begin{proof}
We already know from proposition \ref{ibarrecor} that $\overline{I}$ has a
primary decomposition of the form: 
$$
\overline{I} = \bigcap_{i \in E} q_{i}
$$
with $q_{i}$ homogeneous $m_{\overline{k},P_{i}}$-primary ideal and $\{P_{i}| \ i\in E\}$ is the set of points defined by $\overline{I}$ in $\mathbb{P}^{n}_{\overline{k}}$. From proposition \ref{ibarre}, we have that
\begin{equation}\label{decomp}
I = \bigcap_{i \in E}q_{i} \cap S.
\end{equation}
One has that $q_{i} \cap S$ is a homogeneous $m_{\overline{k},P_{i}}\cap
S$-primary ideal. Thus the primary decomposition of $I$ is deduced from
\eqref{decomp} by discarding those $q_{i} \cap S$ that contain $\bigcap_{j
  \neq i}q_{j} \cap S$ and intersecting those $q_{i} \cap S$ that are
$m_{\overline{k},P_{i_{0}}}\cap S$-primary for the same $P_{i_{0}}$. Firstly,
if $\bigcap_{j \neq i} q_{j} \cap S \subset q_{i} \cap S$, then there exists
a $j_{0} \neq i$ such that $m_{\overline{k},P_{j_{0}}} \cap S =
m_{\overline{k},P_{i}} \cap S$, i.e $P_{j_{0}}$ and $P_{i}$ are
conjugate. Secondly,  $q_{i} \cap S$ and $q_{j} \cap S$ are both
$m_{\overline{k},P_{i_{0}}}\cap S$-primary if and only if $P_{i}, \ P_{j}$
and $P_{i_{0}}$ are conjugate. Thus $I$ has a primary decomposition of the
form  
$$
I = \bigcap_{i \in F}q_{i} \cap S
$$
with $F \subset E$ satisfying that for all $i \in E$ there exists a unique $j
\in F$ such that $P_{i}$ and $P_{j}$ are conjugate. Thus, $(I:u)=I$ if and only
if $u$ does not vanish at any point $P_{j}$ for all $j \in F$, i.e $u$ does
not vanish at any point $P_{i}$ for all $i \in E$. 
\end{proof}
\begin{proposition}\label{point}
 Consider a field extension $k \subset L$ of $k$. Let $\overline{L}$ be the
 algebraic closure of $L$. Let $I$ be a homogeneous ideal in
 $\mathbf{Hilb}^{\mu}_{\mathbb{P}^{n}}(k)$ and let $I_{L} = I
 \otimes_{k} L$. Then, the points defined by $\overline{I_{L}}$ in
 $\mathbb{P}^{n}_{\overline{L}}$ are exactly the image by the field
 extension: 
$$
\overline{k} \subset \overline{L}
$$
of the points defined by $\overline{I}$ in $\mathbb{P}^{n}_{\overline{k}}$.
\end{proposition}
\begin{proof}
The points defined by $\overline{I_{L}}$ in $\mathbb{P}^{n}_{\overline{L}}$ are given by the primary decomposition of $\overline{I_{L}} = I_{L}\otimes_{L}\overline{L}$ in $S \otimes \overline{L}$. We just need to prove that these points are the same as those obtained by the primary decomposition of $\overline{I}$ in $\overline{S}$. In fact, one has that
$$
I_{L} \otimes_{L} \overline{L} = \overline{I} \otimes_{\overline{k}} \overline{L}.
$$
Thus, as 
$$
 \overline{I} = \bigcap_{i \in E} q_{i}
$$
with $q_{i}$ homogeneous $m_{\overline{k},P_{i}}$-primary ideal and $\{P_{i}| \ i\in E\}$ is the set of points defined by $\overline{I}$ in $\mathbb{P}^{n}_{\overline{k}}$; we deduce that $I_{L} \otimes_{L} \overline{L}$ can be written
$$
I_{L} \otimes_{L} \overline{L} = \bigcap_{i \in E} q_{i} \otimes \overline{L}
$$
with $q_{i} \otimes \overline{L}$ homogeneous $m_{\overline{k},P_{i}}\otimes \overline{L}$-primary ideal. But one has that $m_{\overline{k},P_{i}}\otimes \overline{L} = m_{\overline{L},P_{i}}$ ($P_{i}$ considered as a point of $\mathbb{P}^{n}_{\overline{L}}$ via the field inclusion $\overline{k} \subset \overline {L}$). Thus, the points defined by $\overline{I_{L}}$ in $\mathbb{P}^{n}_{\overline{L}}$ are exactly the image by the field extension:
$$
\overline{k} \subset \overline{L}
$$
of $\{P_{i} \in \mathbb{P}^{n}_{\overline{k}}| \ i\in E\}$.
\end{proof}

\bibliographystyle{plain}
\bibliography{paper}

\end{document}